\documentclass[12pt, a4paper]{article}
\pdfoutput=1

\usepackage[height=22cm,width=17cm, centering]{geometry}
\usepackage{setspace}

\usepackage[utf8]{inputenc} 

\usepackage{amsmath}
\usepackage{amssymb, mathrsfs}
\usepackage{graphicx}
\usepackage{subcaption}
\usepackage{color,comment, slashed, bm, textcomp, cancel}
\usepackage[sort&compress, numbers, merge]{natbib}

\definecolor{purple}{rgb}{0.4 ,0, 0.4}
\definecolor{reddish-purple}{rgb}{0.45, 0., 0.35}
\definecolor{bluish-purple}{rgb}{0.35, 0., 0.45}
\usepackage{hyperref}
\hypersetup{colorlinks=true, linkcolor=bluish-purple, citecolor=purple, urlcolor=reddish-purple}

\allowdisplaybreaks

\title{Semianalytic calculation of the gravitational wave spectrum induced by curvature perturbations}
\author{Takahiro Terada}
\date{\today}

\begin{document}

\begin{titlepage}
\begin{center}

\renewcommand{\thefootnote}{\fnsymbol{footnote}} 
\vspace*{22 mm}
Awarded the 20th Seitaro Nakamura Prize\footnote{
This manuscript was newly written for submission to the Prize and is 
largely based on the co-authored work~\cite{Kohri:2018awv}, with the 
exposition reorganized and rewritten. It also includes some additional 
new material. This version contains minor revisions and expanded references compared with the 
manuscript submitted for the 20th Seitaro Nakamura Prize in late April 2025.
}
\vspace{6mm}

{\LARGE Semianalytic calculation of the gravitational wave spectrum induced by curvature perturbations}\\

\vspace{8 mm}
{\Large Takahiro Terada} \\
\vspace{ 3 mm}
\textit{Kobayashi-Maskawa Institute for the Origin of Particles and the Universe, \\ Nagoya University, Tokai National Higher Education and Research System, \\Furo-cho Chikusa-ku, Nagoya 464-8602, Japan} \\

\vspace{10 mm}


\begin{abstract}
    The stochastic gravitational wave (GW) background is secondarily and inevitably induced by the primordial curvature perturbations beyond the first order of the cosmological perturbation theory. We analytically calculate the integration kernel of the power spectrum of the induced GWs, which is the universal part independent of the spectrum of the primordial curvature perturbations, in the radiation-dominated era and in the matter-dominated era. We derive fully analytic expressions of the GW spectrum when possible. As a minor update, we study the case of the top-hat function as the spectrum of the curvature perturbations. We also discuss generalization in the presence of multiple cosmological eras with different equations of state. 
\end{abstract}
\end{center}
\end{titlepage}

\renewcommand{\thefootnote}{\arabic{footnote}}
\setcounter{footnote}{0}
\tableofcontents 
\section{Introduction}

In general, gravitational waves (GWs) are valuable probes of the early Universe and particle physics. Since GWs interact with the gravitational strength, they are hardly absorbed or scattered even in the hot and dense environment where photons cannot go straight. Thus, GWs can convey information about physics that produced themselves in the primordial epoch, whose energy scale could be much higher than what can be probed directly by particle colliders.

Of course, the feeble interactions of the GWs make their detection hard, but the experimental/observational techniques and precisions have become mature enough so that the GWs have been detected directly~\cite{LIGOScientific:2016aoc, LIGOScientific:2018mvr, LIGOScientific:2020ibl, LIGOScientific:2021usb, KAGRA:2021vkt}. Recently, the evidence of Hellings-Downs curve~\cite{Hellings:1983fr}, a smoking-gun signal of the stationary, stochastic, and isotropic GWs, has been found by pulsar timing array (PTA) collaborations~\cite{NANOGrav:2023gor, EPTA:2023fyk, Reardon:2023gzh, Xu:2023wog, InternationalPulsarTimingArray:2023mzf}. While the signal may originate from astrophysical sources such as supermassive black hole binaries~\cite{NANOGrav:2023hfp, NANOGrav:2023tcn, NANOGrav:2023pdq, EPTA:2023gyr, EPTA:2023xxk}, cosmological origins including a first-order phase transition, cosmic strings, domain walls, and enhanced curvature perturbations are also interesting possibilities (see Refs.~\cite{NANOGrav:2023hvm, EPTA:2023xxk, Bian:2023dnv, Figueroa:2023zhu, Ellis:2023oxs} for comparisons). With the various planned GW observatories like SKA~\cite{Carilli:2004nx, Janssen:2014dka, Weltman:2018zrl}, LISA~\cite{LISA:2017pwj, Baker:2019nia}, DECIGO~\cite{Seto:2001qf, Yagi:2011wg, Isoyama:2018rjb, Kawamura:2020pcg}, ET~\cite{Punturo:2010zz, Hild:2010id, Sathyaprakash:2012jk, ET:2019dnz, Abac:2025saz}, and CE~\cite{LIGOScientific:2016wof, Reitze:2019iox}, we are entering the era of GW cosmology.

Among various sources of cosmological GWs, we focus on GWs secondarily induced by primordial curvature perturbations~\cite{10.1143/PTP.37.831, Matarrese:1992rp, Matarrese:1993zf, Matarrese:1997ay, Mollerach:2003nq, Ananda:2006af, Baumann:2007zm}, which are recently called (scalar-)induced GWs (SIGWs) (see  reviews~\cite{Domenech:2021ztg, Inomata:2025wiv} for other early works). There are multiple motivations to study SIGWs. Examples are listed below non-exhaustively.
\begin{itemize}
    \item To probe the primordial curvature perturbations and inflation on small scales~\cite{2010PhRvD..81b3517B, Alabidi:2012ex, Alabidi:2013lya, Orlofsky:2016vbd, Inomata:2018epa, Byrnes:2018txb, Ben-Dayan:2019gll, Inomata:2021zel}, including the effects of non-Gaussianity of the curvature perturbations~\cite{Nakama:2016gzw, Garcia-Bellido:2017aan, Ando:2017veq, Cai:2018dig, Unal:2018yaa, Yuan:2020iwf, Atal:2021jyo, Adshead:2021hnm,  Garcia-Saenz:2022tzu, Abe:2022xur, Li:2023qua, Yuan:2023ofl, Li:2023xtl, Perna:2024ehx, Inui:2024fgk, Iovino:2024sgs, Li:2025met, Zeng:2025cer}. Anisotropy of SIGWs~\cite{Dimastrogiovanni:2022eir, Chen:2022qec, Li:2023qua, Li:2023xtl, Wang:2023ost, Yu:2023jrs, Ruiz:2024weh, Li:2025met, Bodas:2025wef} is interesting in its own right and can, in some cases, serve as a useful probe of non-Gaussianity. SIGWs are sensitive to resonant features from heavy degrees of freedom~\cite{Fumagalli:2020nvq, Fumagalli:2021cel}. They can also probe the metastability of the Electroweak vacuum~\cite{Espinosa:2018eve}. 
    \item To probe the equation of state~\cite{Assadullahi:2009nf, Alabidi:2013lya, Inomata:2019zqy, Inomata:2019ivs, Domenech:2019quo, Hajkarim:2019nbx, Domenech:2020kqm} and sound speed~\cite{Baumann:2007zm, Abe:2020sqb, Balaji:2023ehk} in cosmological epochs and the transition time scale between the epochs~\cite{Inomata:2019zqy, Inomata:2019ivs, Pearce:2023kxp}. SIGWs are also sensitive to the number of relativistic degrees of freedom although it is not specific for the scalar-induced case (see, e.g., Refs.~\cite{Franciolini:2023wjm, Saito:2012bb}). In particular, they can probe the crossover in QCD~\cite{Hajkarim:2019nbx, Abe:2020sqb, Franciolini:2023wjm, Abe:2023yrw} and in models beyond the Standard Model~\cite{Escriva:2023nzn} and probe new physics like supersymmetry~\cite{Saito:2012bb}. Presence of new heavy particles can also be probed via damping of the scalar source~\cite{Yu:2024xmz, Domenech:2025bvr, Yu:2025cqu}. 
    \item To test primordial black hole (PBH) scenarios~\cite{Saito:2008jc, Saito:2009jt, 2010PhRvD..81b3517B, 2011PhRvD..83h3521B, Garcia-Bellido:2017aan, Cai:2018dig, Bartolo:2018evs, Bartolo:2018rku, Unal:2018yaa, Wang:2019kaf}. PBHs can play cosmologically important roles, such as dark matter~\cite{Carr:2016drx, Carr:2020gox, Carr:2020xqk, Escriva:2022duf} and a generator of baryon asymmetry of the Universe (see Refs.~\cite{Bugaev:2001xr, Baumann:2007yr, Fujita:2014hha} and references therein), to name just a few. SIGWs can also be utilized to test some quantum gravity scenarios that involve exotic final states of a black hole~\cite{Domenech:2023mqk, Franciolini:2023osw, Kohri:2024qpd, Balaji:2024hpu, Barman:2024iht, Bhaumik:2024qzd}.
    \item To explain the detected PTA signals, \textit{i.e.}, GWs around nanohertz frequencies~\cite{Madge:2023dxc, Franciolini:2023pbf, Cai:2023dls, Inomata:2023zup, Wang:2023ost, Liu:2023ymk, Abe:2023yrw, Wang:2023sij, Firouzjahi:2023lzg,  Bari:2023rcw, HosseiniMansoori:2023mqh, Balaji:2023ehk, Zhu:2023gmx, Liu:2023pau, Yi:2023tdk, Frosina:2023nxu, Choudhury:2023hfm, Kawasaki:2023rfx, Yi:2023npi, Harigaya:2023pmw, Inomata:2023drn, Liu:2023hpw, Domenech:2024rks}. See also earlier works~\cite{Cai:2019elf,  Chen:2019xse, Vaskonen:2020lbd, DeLuca:2020agl, Kohri:2020qqd, Domenech:2020ers, Inomata:2020xad, Kawasaki:2021ycf, Dandoy:2023jot}. The PTA constraints on PBHs are also discussed in these references as well as Ref.~\cite{Iovino:2024tyg}. See Ref.~\cite{Cecchini:2025oks} for a forecast on future PTAs given the present evidence of nanohertz GWs. 
    \item To test the General Relativity and modified gravity in the primordial epoch beyond the linearized order. See, \textit{e.g.}, Refs.~\cite{Chen:2021nio, Kawai:2021edk, Zhang:2021rqs, Zhang:2022xmm, Yi:2022anu, Feng:2023veu, Tzerefos:2023mpe, Garcia-Saenz:2023zue, Zhang:2023scq, Zhang:2024vfw, Feng:2024yic, Domenech:2024drm, Zhou:2024doz, Kugarajh:2025rbt, Lopez:2025gfu, Zhang:2025mps, Jiang:2025ysb}.\footnote{
    Note that some analyses in the literature adopt the integration formula derived in General Relativity. Its applicability in modified-gravity setups is model-dependent.
    }
\end{itemize}
Thus, there are strong physics cases for SIGWs.

The properties of SIGWs crucially depend on those of the primordial curvature perturbations. For example, the power spectrum of SIGWs depends on the power spectrum of the primordial curvature perturbations. The latter has functional degrees of freedom, so it can be a serious source of uncertainty. As we will see below in detail, the spectrum of SIGWs is given by a convolution integral of an integration kernel, which itself is given by a time integral of an oscillating function, and two instances of the power spectrum of the curvature perturbations.  In model reconstruction or parameter estimation, one typically needs to numerically calculate the spectrum of SIGWs many times, varying underlying parameters, which will be time-consuming.  This is true both for simulations for prospects and for actual data analyses.  Therefore, it is highly beneficial to give an analytic formula for the integration kernel and approximate or exact (semi)analytic formulas for the fully integrated SIGW spectra for typical power spectra of curvature perturbations that are widely used as benchmarks.

In this paper, we analytically compute the integration kernel of SIGWs, which is the main point of this work (or Ref.~\cite{Kohri:2018awv}). It is a universal result applicable to an arbitrary power spectrum of primordial curvature perturbations.  We also give formulas of SIGWs, which are fully analytic when possible, for several example power spectra of curvature perturbations.  As a minor update from Ref.~\cite{Kohri:2018awv}, we add new approximate formulas and exact analytic formulas of SIGWs induced in a radiation-dominated (RD) era and a matter-dominated (MD) era, respectively, by the curvature perturbations whose power spectrum has the top-hat shape.  References of this manuscript include not only the literature at the time of writing Ref.~\cite{Kohri:2018awv} but also later developments. 

Another topic discussed in Ref.~\cite{Kohri:2018awv} is the effect of transitions between an RD era and an MD era.  Since this part was significantly updated~\cite{Inomata:2019zqy, Inomata:2019ivs} and a missing contribution was found in Ref.~\cite{Inomata:2019ivs} after Ref.~\cite{Kohri:2018awv}, the relevance of the naive prescription for the transitions between cosmic eras (in particular, from an MD era to an RD era) in Ref.~\cite{Kohri:2018awv} is limited today. While we do not go into details of the updated work, we give an overview of the effects of transitions between cosmic eras. 

The structure of the paper is as follows. In Sec.~\ref{sec:basics}, we review the formulation of SIGWs.  (Semi)analytic calculations of integrals for SIGWs are performed in Sec.~\ref{sec:analytic}.  We delineate the way to extend the results to a richer cosmological history involving transitions between cosmological epochs with different equations of state in Sec.~\ref{sec:transitions}. We conclude in Sec.~\ref{sec:conclusion}.  Appendix~\ref{sec:region_top-hat} summarizes the integration region for the SIGW spectrum in the case of the top-hat function as the power spectrum of the primordial curvature perturbations. In Appendix~\ref{sec:fits}, we provide several approximate fitting formulas for the SIGW spectrum and its spectral index. Throughout the paper, we use the natural unit where $c$, $\hbar$, $k_\text{B}$, and $8\pi G = 1/M_\text{P}^2$ are set to unity unless we emphasize the dependence.  

\section{Basics of the induced gravitational waves \label{sec:basics}}
We consider perturbations around the Friedmann-Lema\^itre-Robertson-Walker spacetime, whose invariant interval in the Newtonian gauge is
\begin{align}
    \mathrm{d}s^2 = - a^2 (1 + 2 \Phi) \mathrm{d}\eta^2 + a^2 \left((1 - 2 \Psi)\delta_{ij} + \frac{1}{2} h_{ij} \right) \mathrm{d}x^i \mathrm{d}x^j,
\end{align}
where $\eta$ is the conformal time, $a(\eta)$ is the scale factor, $\Phi$ and $\Psi$ are the first-order scalar perturbations corresponding to the gravitational potential and the curvature perturbations, and $h_{ij}$ is the second-order tensor perturbations. Here, we are interested in the second-order tensor mode induced by the first-order scalar modes at the second order of the cosmological perturbation theory,\footnote{
The third-order induced GWs were studied in Refs.~\cite{Yuan:2019udt, Zhou:2021vcw, Chang:2022nzu, Wang:2023sij}. Higher-order effects were discussed in Ref.~\cite{Zhou:2024ncc}.
} so we neglected the first-order tensor mode,\footnote{
If the first-order tensor mode is not negligible, its interference with the third-order tensor mode (schematically, $\langle h^{(1)}h^{(3)}\rangle$ where $(i)$ denotes the $i$-th order of cosmological perturbation) gives the contribution to the power spectrum of the GWs at the same order with the contribution of our interest (schematically, $\langle h^{(2)} h^{(2)}\rangle $)~\cite{Chen:2022dah}.
} the vector mode, and irrelevant higher-order modes.\footnote{
For GWs induced not only by scalar modes but also by vector and tensor modes, see Ref.~\cite{Gong:2019mui}. 
} By the same token, we assume the absence of the anisotropic stress at the first order, which implies $\Phi = \Psi$.\footnote{
An analysis without this assumption is given in Ref.~\cite{Baumann:2007zm}.
}  Here and in what follows, we basically follow the convention/notation of Ref.~\cite{Inomata:2016rbd, Kohri:2018awv}. See also Refs.~\cite{Ananda:2006af, Baumann:2007zm} for derivation. 

The tensor field is decomposed into its Fourier components 
\begin{align}
    h_{ij}(\eta , \bm{x}) =& \int \frac{\mathrm{d}^3 k}{(2\pi)^{3/2}} \sum_{\lambda = +, \times} e^{\lambda}_{ij} (\bm{k}) h_{\bm{k}}^\lambda (\eta) e^{i \bm{k} \cdot \bm{x}},
\end{align}
where $\lambda = +, \times$ denotes the polarization mode, $e^\lambda_{ij}$ is the polarization tensor: $e^+_{ij}(\bm{k}) = (e_i(\bm{k})e_j(\bm{k}) - \bar{e}_i (\bm{k}) \bar{e}_j (\bm{k}))/\sqrt{2}$ and $e^\times_{ij}(\bm{k})= (e_i(\bm{k})\bar{e}_j(\bm{k}) + \bar{e}_i (\bm{k})e_j(\bm{k}))/\sqrt{2}$ with $e_i(\bm{k})$ and $\bar{e}_i(\bm{k})$ denoting two mutually orthonormal polarization vectors.  

The (second-order) equation of motion for $h_{\bm{k}}^\lambda$ is as follows
\begin{align}
    h_{\bm{k}}^\lambda{}''(\eta) + 2 \mathcal{H}(\eta) h_{\bm{k}}^\lambda{}'(\eta) + k^2 h_{\bm{k}}^\lambda(\eta) = 4 S_{\bm{k}}^{\lambda}(\eta),
\end{align}
where a prime denotes the conformal time derivative, $\mathcal{H} = a'/a$ is the conformal Hubble parameter, and $S_{\bm{k}}^{\lambda}(\eta)$ is the source term. This coupling is a built-in effect in General Relativity with the Einstein-Hilbert term when expanded in terms of the perturbation fields. The source term has the following expression
\begin{align}
    S_{\bm{k}}^{\lambda} = \int \frac{\mathrm{d}^3 q}{(2\pi)^{3/2}} e_{ij}^{\lambda}(\bm{k}) q_i q_j \left( 2 \Phi_{\bm{q}} \Phi_{\bm{k}-\bm{q}} + \frac{4}{3 ( 1 + w)}\left( \mathcal{H}^{-1}\Phi'_{\bm{q}} + \Phi_{\bm{q}} \right) \left( \mathcal{H}^{-1}\Phi'_{\bm{k}-\bm{q}} + \Phi_{\bm{k}-\bm{q}} \right)  \right),
\end{align}
where we used $- 2 \dot{H} = 3 (1 + w)H^2$ with a dot denoting time derivative, and $w = P/\rho$ is the equation-of-state parameter with $P$ and $\rho$ denoting the pressure and the energy density of the cosmological fluid. 

The tensor field $h_{\bm{k}}^{\lambda}(\eta)$ can be formally solved by the Green's function method\footnote{
For a study using the in-in formalism, see Ref.~\cite{Ota:2022xni}. 
} as follows
\begin{align}
    a(\eta) h_{\bm{k}}^\lambda (\eta) = 4 \int_0^\eta \mathrm{d}\bar \eta G_{k}(\eta, \bar \eta) a(\bar \eta) S_{\bm{k}}^\lambda (\bar \eta),
\end{align}
where $G_k (\eta, \bar \eta)$ is Green's function satisfying
\begin{align}
    G''_k (\eta, \bar \eta) + \left( k^2 - \frac{a''(\eta)}{a(\eta}\right) G_k (\eta, \bar \eta) = \delta (\eta - \bar \eta ).
\end{align}
Therefore, if we know the time dependence of $S_{\bm{k}}^{\lambda}(\eta)$, then we formally know the time dependence of $h_{\bm{k}}^{\lambda}(\eta)$. 

The time dependence of the source term, of course, depends on that of $\Phi_{\bm q}(\eta)$, which we now discuss.  In a general background specified by the sound speed $c_\text{s}$ and possible nonadiabatic pressure $\delta P_\text{nad}$,\footnote{The pressure perturbation is generally decomposed as $\delta P = c_\text{s}^2 \delta \rho + \delta P_{\text{nad}}$.} the equation of $\Phi$ reads (see, \textit{e.g.}, Ref.~\cite{Mukhanov:2005sc})
\begin{align}
    \Phi''_{\bm{k}} + 3 \mathcal{H} ( 1 + c_\text{s}^2 ) \Phi'_{\bm{k}} + \left(2 \mathcal{H}' + (1 + 3 c_\text{s}^2) \mathcal{H}^2 + c_\text{s}^2 k^2 \right) \Phi_{\bm{k}} = \frac{a^2}{2} \delta P_{\text{nad}}. \label{EoM_Phi_complete}
\end{align}
In this work, we primarily focus on the adiabatic ($\delta P_\text{nad} = 0$) and barotropic ($P \propto \rho$) case, for which $c_\text{s}^2 = w$.\footnote{
GWs induced by isocurvature perturbations were studied in Refs.~\cite{Domenech:2021and, Domenech:2023jve}. 
} 
In this case, the equation simplifies to 
\begin{align}
    \Phi''_{\bm{k}}(\eta) + 3 (1 + w) \mathcal{H} \Phi'_{\bm{k}}(\eta) + w k^2 \Phi_{\bm{k}}(\eta) = 0. \label{EoM_Phi}
\end{align}
Since $\mathcal{H}\sim 1/\eta$, the time dependence typically depends on the combination $k \eta$ up to the normalization for each mode, which is given by the initial condition. Introducing the transfer function $T_\Phi (k \eta) $, we can express $\Phi_{\bm{k}}(\eta) = T_\Phi (k \eta) \Phi_{\bm{k}}(0)$. The primordial value $\Phi_{\bm{k}}(0)$ is related to the primordial curvature perturbation on the uniform-density gauge $\zeta_{\bm{k}}$ by $\Phi_{\bm{k}} = - \frac{3+3w}{5+3w} \zeta_{\bm{k}}$.

Substituting $\Phi_{\bm{k}}(\eta) = T_\Phi (k \eta) \Phi_{\bm{k}}(0)$ to the source term $S_{\bm{k}}^{\lambda}(\eta)$, one can compute the power spectrum of the tensor perturbations, $\mathcal{P}_h(\eta, k)$, which is defined by
\begin{align}
    \langle h_{\bm{k}}^{\lambda}(\eta) h_{\bm{k}'}^{\lambda'}(\eta)\rangle = \delta^{\lambda\lambda'}\delta^3(\bm{k}+\bm{k}') \frac{2\pi^2}{k^3} \mathcal{P}_h(\eta, k).
\end{align}
In this computation, one needs to evaluate the four-point correlation function of the primordial curvature perturbations. In this work, we assume the Gaussian statistics of perturbations.  For the discussions on the effects of non-Gaussianity, see Refs.~\cite{Nakama:2016gzw, Garcia-Bellido:2017aan, Ando:2017veq, Cai:2018dig, Unal:2018yaa, Yuan:2020iwf, Atal:2021jyo, Adshead:2021hnm,  Garcia-Saenz:2022tzu, Abe:2022xur, Li:2023qua, Yuan:2023ofl, Li:2023xtl, Perna:2024ehx, Inui:2024fgk, Iovino:2024sgs, Li:2025met, Zeng:2025cer}. 
After some algebra~\cite{Inomata:2016rbd}, it is given by
\begin{align}
    \mathcal{P}_h(\eta, k) = 4 \int_0^\infty \mathrm{d}u \int_{|u-1|}^{u+1} \mathrm{d}v \left(\frac{1-2(u^2+v^2)-4u^2 v^2 + u^4 + v^4}{4 u v}\right)^2 I^2 (u, v, k \eta) \mathcal{P}_\zeta(u k) \mathcal{P}_\zeta (v k), \label{P_h}
\end{align}
where the integration variables $u$ and $v$ originates from the wave-number integrals, $u = |\bm{k}-\widetilde{\bm{k}}|/k$ and $v = \widetilde{k}/k$ with $\widetilde{\bm{k}}$ denoting the integrated wave number of one of the two scalar modes. The integration kernel $I(u,v,k \eta)$ is defined as follows 
\begin{align}
    I(u, v, k \eta) = \int_0^{k\eta} \mathrm{d}(k \bar{\eta}) \frac{a(\bar \eta)}{a(\eta)} k G_k(\eta, \bar \eta) f(u, v, k \bar \eta),
\end{align}
where $f(u, v, k \bar \eta)$ is defined as 
\begin{align}
    f(u, v, k\bar \eta) =& \left(\frac{3+3w}{5+3w}\right)^2 \left[ \frac{2(3w+5)}{3(1+w)} T_\Phi( uk\bar \eta) T_\Phi(v k \bar \eta) +  \frac{2(1+3w)}{3(1+w)}\left(uk\bar \eta T'_\Phi(u k \bar \eta) T_\Phi(v k \bar \eta) \right.\right. \nonumber \\
    & \left. \left. + v k \bar \eta T_\Phi(u k \bar \eta) T'_\Phi(v k \bar \eta) \right) + \frac{(1+3w)^2}{3(1+w)} u v (k\bar\eta)^2 T'_\Phi(u k \bar \eta)T'_\Phi(v k \bar \eta) \right],
\end{align}
where we used $\mathcal{H}= 2/((1+3w)\eta)$, and a prime here denotes the differentiation with respect to the argument.

Let us explain the above formulas. First, $\mathcal{P}_h$ depends on the quadratic form of $\mathcal{P}_\zeta$ since we are discussing the second-order SIGWs. The function $I(u, v, k \eta)$ contains all the information of the dynamics as $\eta$-dependence only appears in it on the right-hand side of eq.~\eqref{P_h}. In the definition of $I(u, v, k \eta)$, the dynamics of the source term is described in the function $f(u, v, k \bar\eta)$, while $k G_k(\eta, \bar \eta)$ describes the time evolution of each GW mode from the production time $\bar \eta $ to the evaluation time $\eta$. The factor $a(\bar \eta)/a(\eta)$ represents the redshift of the GWs. Coming back to eq.~\eqref{P_h}, the kinematic factor dependent on $u$ and $v$ comes from the contraction between the polarization tensor and the wave numbers of the scalar source modes. Finally, the restriction on the integration region accounts for the momentum conservation. Note that the above expressions, including the integration domain, are symmetric under the exchange of $u$ and $v$~\cite{Inomata:2016rbd}. 

For the practical purpose of numerical integration, we introduce another representation of eq.~\eqref{P_h} by the changes of variables:
\begin{align}
    &\begin{cases}
        u = (t - s + 1)/2, \\
        v = (t + s + 1)/2,
    \end{cases}
    & &\text{or} & &
    \begin{cases}
        t = u + v - 1,\\
        s = v - u.
    \end{cases}
\end{align}
The alternative expression is~\cite{Kohri:2018awv} 
\begin{align}
    \mathcal{P}_h(\eta, k) = 4 \int_0^\infty \mathrm{d}t\int_0^1 \mathrm{d}s \left(\frac{t(t+2)(s^2-1)}{(t+s+1)(t-s+1)} \right)^2 I(u, v, k \eta)^2 \mathcal{P}_\zeta(u k) \mathcal{P}_\zeta(v k),
\end{align}
where $u = (t-s+1)/2$ and $v=(t+s+1)/2$. An advantage of this expression is the simplification of the integration domain. 

Finally, let us relate these quantities to observables. In cosmology, the frequency-dependent intensity of GWs is often parametrized by $\Omega_\text{GW}(\eta, f) =\rho_\text{GW}(\eta, f)/\rho_\text{total}$, where $\rho_\text{total} = 3 H^2 M_\text{P}^2$ is the total energy density.\footnote{The total energy density of the GWs is obtained by the integral $\rho_\text{GW}(\eta) = \int \mathrm{d}\ln (f/f_*) \, \rho_\text{GW}(\eta, f)$, with $f_*$ being an arbitrary frequency to make the argument of the logarithm dimensionless.} Talking about observables, we are interested in GW modes on subhorizon scales, where GW modes oscillate rapidly so that General Relativistic effects relevant around the horizon scale and gauge ambiguity\footnote{
There were discussions on the gauge (in)dependence of the induced GWs~\cite{Hwang:2017oxa, Gong:2019mui, Tomikawa:2019tvi,  DeLuca:2019ufz, Inomata:2019yww, Yuan:2019fwv, Lu:2020diy, Ali:2020sfw, Chang:2020iji, Chang:2020mky, Domenech:2020xin, Gurian:2021rfv,  Cai:2021jbi, Cai:2021ndu, Ota:2021fdv,  Ali:2023moi, Comeau:2023mxi, Yuan:2024qfz, Yuan:2025seu}.  For the practical purpose, the calculation in the Newtonian gauge (equivalent to the synchronous gauge well after the horizon entry~\cite{DeLuca:2019ufz, Inomata:2019yww, Yuan:2019fwv}) is physical and simplest at least in an RD era. The gauge dependence issue for the tensor modes in an MD era is less clear. 
} is practically negligible.\footnote{\label{fn:MLambda}
For large-scale tensor modes induced after the matter-radiation equality, this assumption is not valid. See Ref.~\cite{Sipp:2022kmb}.
}  In this situation, $\rho_\text{GW}$ can be thought of as a sum of the kinetic and potential energy densities, and it can be obtained by an oscillation average of either the kinetic or potential energy density. Specifically, the second-order graviton action in our convention is $S= (M_\text{P}^2/32)\int \mathrm{d}\eta \mathrm{d}^3 x a^2 (h'_{ij}h'_{ij} - h_{ij,k}h_{ij,k})$, so we can define the energy density of the second-order induced GWs as $\rho_\text{GW}= (M_\text{P}^2/(16a^2))\langle \overline{h_{ij, k}h_{ij,k}}\rangle$, where the overline denotes the oscillation average. Then, $\Omega_\text{GW}(\eta, f)$ is given by
\begin{align}
    \Omega_\text{GW}(\eta, f) = \frac{1}{24} \left(\frac{k}{\mathcal{H}(\eta)} \right)^2 \overline{\mathcal{P}_h (\eta, k)}, \label{Omega_GW_def}
\end{align}
where the frequency $f$ and the wave number $k$ are related to each other as usual by $2 \pi f = k$. 
We have added the contributions from both polarization modes $\lambda = +, \times$. 

Suppose that the GWs are induced during or before the (latest) RD era. Since GWs behave like radiation, $\Omega_\text{GW}$ becomes a constant at $\eta = \eta_\text{c}$ during the RD era up to the change of numbers of relativistic degrees of freedom.  Once we derive its value, it is related to the present value via 
\begin{align}
    \Omega_\text{GW}(\eta_0, f) = \Omega_{\text{r},0} \left(\frac{g_{*}(T_\text{c})}{g_{*}(T_0)}\right)\left(\frac{g_{*s}(T_0)}{g_{*s}(T_\text{c})}\right)^{4/3} \Omega_\text{GW}(\eta_\text{c}, f), \label{Omega_GW_obs}
\end{align}
where the subscripts $0$ and c denote the present time and $\eta= \eta_\text{c}$, respectively, $g_{*}(T)$ and $g_{*s}(T)$ are the effective number of relativistic degrees of freedom at the temperature $T$ for the energy density and the entropy density, respectively.\footnote{We have used $\rho_\text{r}\propto g_*(T)T^4$, $\rho_\text{GW}(\eta, k) \propto a^{-4}$, and the adiabatic expansion $g_{*s}(T)a^3 T^3 = \text{const.}$ Precisely speaking, the number of degrees of freedom may change in the time integral of $I(u, v, k \eta)$, but we neglect this dependence, assuming that the production time of GWs is dominated around some time. }  In the following discussion, we are mostly interested in $\Omega_\text{GW}(\eta_\text{c}, f)$ since the rest of the factors are approximately constant common factors.

Having introduced various definitions, let us recap the motivation for analytically computing the integral $I(u, v, k \eta)$. For this purpose, let us consider the standard case with the RD era. As we will see below, the function $f(u,v,k\bar\eta)$ is an oscillating function of $k\bar \eta$, and $G_k (\eta, \bar \eta)$ is also an oscillating function of $k(\eta - \bar \eta)$. They start oscillations when the mode enters the Hubble horizon, and the oscillations become extremely rapid relative to the Hubble time scale at late times.  It is doable but computationally expensive to perform such an integral with respect to $k \bar \eta$ numerically. Moreover, one has to redo the integral for each choice of $u$ and $v$.  Alternatively, one may choose to integrate over $u$ and $v$ first, but in this case, one cannot reuse the result when one considers different choices of $\mathcal{P}_\zeta (k)$.  Thus, it is beneficial if the integral can be performed analytically once and for all.  Once we have an analytic expression for $I(u, v, k \eta)$, we can easily obtain the oscillation average $\overline{I(u, v, k)^2}$, which is no longer a rapidly oscillating function of $k$.  The resulting two-dimensional integral over $u$ and $v$ (or equivalently, $t$ and $s$) depends on $\mathcal{P}_\zeta (k)$, but it is relatively simple (without a rapidly oscillating function). In the next section, we analytically compute the integral $I(u,v,k \eta)$. A partial calculation was done in Ref.~\cite{Ananda:2006af}, and we complete the calculation to obtain a compact formula. 

\section{Calculation of the induced gravitational waves \label{sec:analytic}}
In this section, we consider GWs induced during a pure RD universe in subsection~\ref{ssec:RD} and during a pure MD universe in subsection~\ref{ssec:MD}. 
\subsection{Radiation-dominated universe\label{ssec:RD}}
In an RD era, the equation-of-state parameter is $w = 1/3$, the scale factor behaves as $a \propto \eta$, and the conformal Hubble parameter satisfies $\mathcal{H}= 1/\eta$. The equation of motion for $T_\Phi$ becomes
   $ T''_\Phi + \frac{4}{\eta}T'_\Phi + \frac{k^2}{3} T_\Phi = 0$. 
The normalized and regular ($T_\Phi (0) = 1$) solution is given by the spherical Bessel function of the first kind $3 \sqrt{3} j_1 (x/\sqrt{3})/x$, whose explicit form is 
\begin{align}
    T_\Phi (x) = \frac{9}{x^2}\left( \frac{\sin (x/\sqrt{3})}{x/\sqrt{3}} - \cos (x/\sqrt{3}) \right),
\end{align}
where $x \equiv k \eta$ is introduced for compact notation. The source function $f(u,v,\bar{x})$ becomes
\begin{align}
    f_\text{RD}(u, v, \bar{x}) =& \frac{12}{u^3 v^3 \bar{x}^6} \left( 18 u v \bar{x}^2 \cos \frac{u \bar{x}}{\sqrt{3}} \cos \frac{v \bar{x}}{\sqrt{3}} + \left( 54 - 6 (u^2 + v^2) \bar{x}^2 + u^2 v^2 \bar{x}^4 \right) \sin \frac{u \bar{x}}{\sqrt{3}} \sin \frac{v \bar{x}}{\sqrt{3}} \right. \nonumber \\
    & \left. + 2 \sqrt{3} u \bar{x} (v^2 \bar{x}^2 - 9) \cos \frac{u \bar{x}}{\sqrt{3}} \sin \frac{v \bar{x}}{\sqrt{3}} + 2 \sqrt{3} v \bar{x} ( u^2 \bar{x}^2 - 9) \sin \frac{u \bar{x}}{\sqrt{3}} \cos \frac{v \bar{x}}{\sqrt{3}} \right),
\end{align}
where $\bar x \equiv k \bar \eta$. 
This is equal to $4/3$ at $\bar{x} = 0$ and decays as $\sim 12/(u v \bar x^2)$ at $\bar x\gg 1$. 

Green's function for GWs satisfy $G''_k (\eta, \bar \eta) + k^2 G_k (\eta, \bar \eta) = \delta (\eta -\bar \eta)$ in the RD era. The retarded solution is
\begin{align}
    k G_k(\eta, \bar \eta) = \sin (x - \bar{x}).
\end{align}

Combining the above formulas, one can derive the analytic formula of the integration kernel $I(u, v, x)$. To this end, we repeatedly use the trigonometric addition theorem and integration by parts~\cite{Ananda:2006af}. 
After a straightforward calculation, we obtain
\begin{align}
   x I_\text{RD}(u, v, x) =& \frac{3}{4u^3 v^3} \left\{ -\frac{4}{x^3} \left( u v (u^2 + v^2 -3)x^3 \sin x - 6 u v x^2 \cos \frac{u x}{\sqrt{3}} \cos \frac{v x}{\sqrt{3}} \right. \right. \nonumber \\
   & + 6 \sqrt{3} u x \cos \frac{u x}{\sqrt{3}} \sin \frac{v x}{\sqrt{3}} + 6\sqrt{3} v x  \sin \frac{u x}{\sqrt{3}} \cos \frac{v x}{\sqrt{3}} \nonumber \\
   & \left. - 3 (6 + (u^2 +v^2 -3)x^2) \sin \frac{u x}{\sqrt{3}} \sin \frac{v x}{\sqrt{3}} \right)  \nonumber \\
   & + (u^2 +v^2 -3)^2 \left[ \sin x \left( \mathrm{Ci}\left( \left(1- \frac{v-u}{\sqrt{3}} \right)x \right) + \mathrm{Ci}\left( \left( 1 + \frac{v-u}{\sqrt{3}}\right)x\right) \right.\right. \nonumber \\
   & \left.- \mathrm{Ci}\left( \left| 1 - \frac{u+v}{\sqrt{3}}\right|x \right)- \mathrm{Ci}\left(\left(1 + \frac{u+v}{\sqrt{3}}\right) x\right) + \log \left| \frac{(u+v)^2-3}{(u-v)^2-3}\right|\right) \nonumber \\
   & + \cos x \left( - \mathrm{Si}\left(\left(1 - \frac{v-u}{\sqrt{3}}\right)x\right) - \mathrm{Si}\left(\left(1+\frac{v-u}{\sqrt{3}}\right)x\right) \right. \nonumber \\
   & \left.\left. \left. + \mathrm{Si}\left(\left(1 - \frac{u+v}{\sqrt{3}}\right)x\right) + \mathrm{Si}\left(\left(1+\frac{u+v}{\sqrt{3}}\right)x\right) \right) \right ] \right\}, \label{I_RD}
\end{align}
where $\mathrm{Si}$ and $\mathrm{Ci}$ are defined as
\begin{align}
    \mathrm{Si}(x) =& \int_0^x \mathrm{d}\bar{x} \frac{\sin \bar{x}}{\bar x}, & 
    \mathrm{Ci}(x) =& - \int_x^\infty \mathrm{d}\bar x \frac{\cos \bar x}{\bar x}.
\end{align}
We have also used the fact 
\begin{align}
    \int_0^x \mathrm{d}\bar x \frac{\cos ( A \bar x )- \cos ( B \bar x)}{\bar x} = \mathrm{Ci}(A x) - \log (A x ) - \mathrm{Ci}(B x) + \log (B x),
\end{align}
with $A$ and $B$ being coefficients. 

For $x\ll 1$, $I_\text{RD}(u, v, x)$ rises as $x^2/2$, while for $x \gg 1$, it oscillates with the amplitude decaying as $1/x$ whose coefficient depend on $u$ and $v$. 

In the late-time limit, $x \gg 1$, it reduces to
\begin{align}
    x I_\text{RD}(u, v, x \gg 1)] =& \frac{3 (u^2 + v^2 -3)}{4 u^3 v^3} \left( \left( -4u v + (u^2 + v^2 -3) \log \left| \frac{3-(u+v)^2}{3-(u-v)^2} \right|\right) \sin x \right. \nonumber \\
    & \left. - \pi (u^2 + v^2 -3) \Theta (u + v - \sqrt{3}) \cos x \right),
\end{align}
where $\Theta$ is the Heaviside step function. We have used $\lim_{x\to \pm \infty}\mathrm{Si}(x) = \pm \pi/2$ and $\lim_{x\to +\infty}\mathrm{Ci}(x) = 0$. 
As expected, it oscillates sinusoidally. Taking the oscillation average, we finally obtain 
\begin{align}
    x^2 \overline{I_\text{RD}^2(u, v, x\gg 1)} =& \frac{1}{2} \left(\frac{3(u^2+v^2-3)}{4 u^3 v^3}\right)^2 \left( \left( -4u v + (u^2 + v^2 -3) \log \left| \frac{3-(u+v)^2}{3-(u-v)^2} \right|\right)^2 \right. \nonumber \\
    & \left. + \pi^2 (u^2 +v^2 -3)^2 \Theta (u + v - \sqrt{3}) \right). \label{I_RD_osc_avg}
\end{align}
In terms of the variables $t$ and $s$, it is
\begin{align}
    x^2 \overline{I_\text{RD}^2(t,s,x\gg1)} = & \frac{288 (t^2 + 2t +s^2 -5)^2}{(t+s+1)^6 (t-s+1)^6} \left( \frac{\pi^2}{4}\left( t^2 + 2 t + s^2 -5 \right)^2 \Theta (t - (\sqrt{3}-1)) \right. \nonumber \\
    & \left. + \left( \! (t+s+1)(t-s+1) - \! \frac{1}{2}(t^2 + 2 t + s^2 -5) \log \left| \frac{t^2 + 2 t -2}{3-s^2} \right| \right)^{\! 2} \right)\!. \label{I_RD_osc_avg_alt}
\end{align}
These formulas are our main results.\footnote{
For the purpose of an analytic study, it may be useful to have an expression without the absolute value. Eq.~\eqref{I_RD_osc_avg} can be rewritten as follows
\begin{align}
    x^2 \overline{I_\text{RD}^2(u, v, x\gg 1)} =& \frac{1}{2} \left(\frac{3(u^2+v^2-3)}{4 u^3 v^3}\right)^2  \mathrm{Re} \left[ \left((u^2 + v^2 -3) \log \left( \frac{(u+v)^2-3}{3-(u-v)^2}  \right)- 4 u v \right)^2 + \pi^2 (u^2 +v^2 -3)^2  \right]. \label{I_RD_osc_avg_ana}
\end{align}
A similar expression in terms of $t$ and $s$ is also possible. 
}
Similar results were obtained in Ref.~\cite{Espinosa:2018eve}.\footnote{
The formulas derived in Ref.~\cite{Espinosa:2018eve}, which appeared one arXiv day before Ref.~\cite{Kohri:2018awv}, correspond to the contribution induced after the horizon entry of the tensor mode $\bar x \geq 1$. The comparison between the results is given in Appendix D of Ref.~\cite{Espinosa:2018eve} after our private communication. Except for the time integral region, these results are fully consistent with each other after taking into account the different conventions and notations. 
}

Let us consider some examples.

\subsubsection{Example 1: Delta function}
First, consider the delta-function case for the power spectrum of the primordial curvature perturbations
\begin{align}
    \mathcal{P}_\zeta (k) = A \delta (\log(k/k_*)), \label{P_zeta_delta}
\end{align}
where $A$ is the overall normalization and $k_*$ is the wave number of the peak. The technical virtue of the delta-function case is, of course, that the integral becomes trivial.  On the other hand, the delta-function is a rather rough approximation of a sharp peak, though it is often considered in the literature, \textit{e.g.}, in the context of an approximately monochromatic PBH formation scenario. We will shortly come back to the limitation of the delta-function approximation. For the PBH application to dark matter in the asteroid mass range, $k_*$ should be taken around $\mathcal{O}(10^{12} \sim 10^{14}) \, \mathrm{Mpc}^{-1}$. 

\begin{figure}[tbhp]
\begin{center}
\includegraphics[width=0.6 \columnwidth]{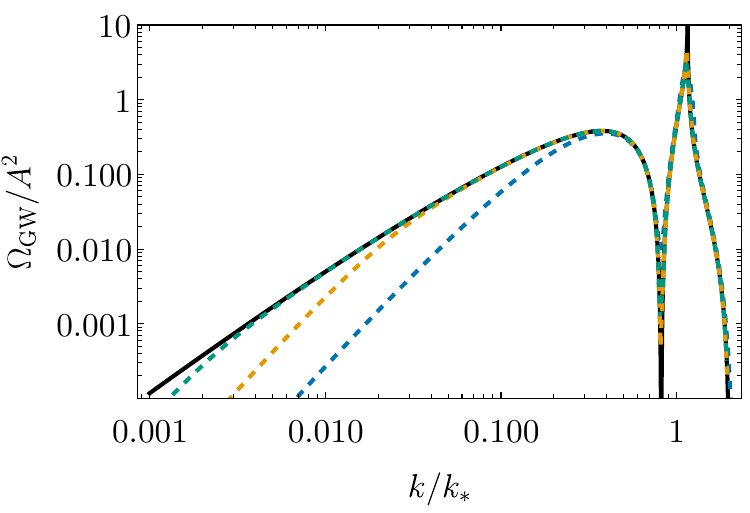}
\end{center}
\caption{GW spectrum induced by the delta-function $\mathcal{P}_\zeta (k)$ during the RD era (solid black line). Also shown in dashed blue, orange, and bluish-green lines for comparison are those induced from the top-hat function with $\Delta = 10^{-1}$, $10^{-2}$, and $10^{-3}$, respectively. For the top-hat function case, $k_*$ should be regarded as $k_\text{med}\equiv \sqrt{k_\text{min}k_\text{max}}$. }
\label{fig:Omega_GW_RD_delta}
\end{figure}

The SIGW spectrum is 
\begin{align}
    \Omega_\text{GW}(\eta_\text{c}, k) =& \frac{3 A^2}{64} \kappa^2 (3 \kappa^2 -2)^2 \left( \frac{4 - \kappa^2}{4} \right)^2 \Theta (2 - \kappa) \nonumber \\
    & \times \left( \pi^2 (3 \kappa^2 -2)^2 \Theta (2 - \sqrt{3} \kappa) + \left( 4 + (3 \kappa^2 - 2) \log \left|1 -  \frac{4}{3 \kappa^2} \right| \right)^2  \right), \label{Omega_GW_RD_delta}
\end{align}
where $\kappa \equiv k / k_*$ is the dimensionless wavenumber. This is plotted in Fig.~\ref{fig:Omega_GW_RD_delta} by the solid black line.  
Since we consider the second-order effect, the maximal wave number of the induced GWs is twice the source wave number, i.e., $\kappa \leq 2$. The peak is at $\kappa = 2/\sqrt{3}$, which satisfies the resonance condition: momentum conservation $\bm{k} = \bm{k}_1 + \bm{k}_2$ and energy conservation $k = (k_1 + k_2)/\sqrt{3}$, where the factor $1/\sqrt{3}$ represents the speed of sound. While generic modes of the GWs are dominantly produced around the horizon reentry, the resonant mode is kept produced on subhorizon scales~\cite{Ananda:2006af}, leading to the logarithmic singular peak in the limit of infinite time $x \to \infty$.  Two comments on this singular behavior are in order.  First, any detector has a finite resolution, and the logarithmic divergence will be smeared at observation. With such an effect, the intensity of GWs is no longer divergent. This can be expected from the fact that the integration of $\Omega_\text{GW}(\eta_\text{c}, k)$ with respect to $\ln \, \kappa$ around the peak $\kappa = 1$ is finite~\cite{Inomata:2016rbd}. Second, it was recently pointed out that the logarithmic peak is smeared by a dissipative effect~\cite{Yu:2024xmz, Domenech:2025bvr, Yu:2025cqu}.  There is a zero at $\kappa = \sqrt{2/3}$.  This feature is not protected when we add corrections such as from non-Gaussianity~\cite{Cai:2018dig} and from the third-order effect~\cite{Wang:2023sij}. 

Another remarkable aspect of the SIGW spectrum for the delta-function case is its infrared (IR) features. Whenever GWs are produced during a finite period in an RD era, their IR power of $\Omega_\text{GW}(f)$ is universally governed by causality and simple statistics, and it scales as $f^3$~\cite{Caprini:2009fx, Cai:2019cdl, Yuan:2019wwo, Hook:2020phx}. On the other hand, the IR power of eq.~\eqref{Omega_GW_RD_delta} is $f^2$.  This is explained, e.g., in Ref.~\cite{Ota:2022xni} by noting that the delta-function power in Fourier space corresponds to infinitely extending waves in position space (rather than wave packets), and in this sense, the initial condition violates causality. For the change of the power-law behavior from a finite-width case to the delta-function limit, see Ref.~\cite{Pi:2020otn}, in which lognormal power spectra are studied. In the perspective of Ref.~\cite{Pi:2020otn}, the transition frequency $f_\text{tr}$ to the causal behavior $\Omega_\text{GW}(f \ll f_\text{tr}) \sim f^3$ vanishes in the delta-function limit, $f_\text{tr} \to 0$, so that there is no $f^3$ regime. In Fig.~\ref{fig:Omega_GW_RD_delta}, we see a similar limiting behavior for narrower and narrower top-hat functions.

Another interesting feature of the IR part is that the $f$ dependence does not obey the pure power law. It involves a logarithmic dependence~\cite{Yuan:2019wwo}. This point is not limited to the delta-function case, but it is a characteristic feature caused by the resonance for GWs induced in the RD era. The dissipative effect mentioned above~\cite{Yu:2024xmz, Domenech:2025bvr, Yu:2025cqu} eliminates this effect for sufficiently low frequencies.

\subsubsection{Example 2: Power law\label{sssec:RD_power-law}}

Next, we consider the power-law spectrum
\begin{align}
    \mathcal{P}_\zeta (k)= A \left(\frac{k}{k_*}\right)^{n_\text{s}-1}, \label{P_zeta_power-law}
\end{align}
where $A$ is an overall normalization, $k_*$ is an arbitrary pivot scale, and $n_\text{s}$ is the scalar spectral index.  Again, this is used as a simple toy model for an illustration purpose, and we do not worry about nonperturbative physics in the regime $\mathcal{P}_\zeta(k)>1$. 

\begin{figure}[tbhp]
\begin{center}
\includegraphics[width=0.6 \columnwidth]{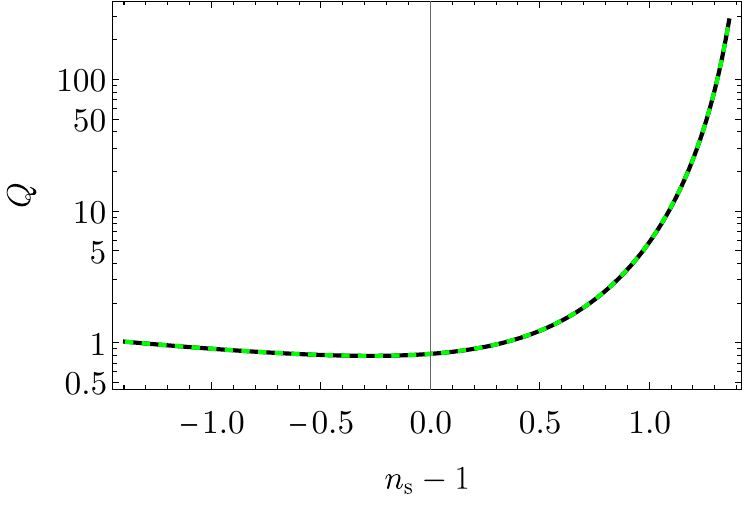}
\end{center}
\caption{Dependence of the numerical coefficient $Q$ on the power $n_\text{s}$ of $\mathcal{P}_\zeta (k) = A (k/k_*)^{n_\text{s}-1}$. The solid black line is the numerical result, while the dashed green line is a fit by the Pad\'e approximation~\eqref{Q_fit}. }
\label{fig:Q}
\end{figure}

Since $\mathcal{P}_\zeta (k)$ is a monomial function of $k$ in this case, the dependence on $k$ factorizes in the formula of $\Omega_\text{GW}(f(k))$. Specifically, $\mathcal{P}_\zeta (u k) \mathcal{P}_\zeta (v k) = (u v)^{n_\text{s}-1} \mathcal{P}_\zeta(k)^2$, and we see, in particular, that $\Omega_\text{GW}(f) \propto \mathcal{P}_\zeta (k)^2$, reflecting the second-order nature of the SIGWs.  
With the help of the analytic formula, eq.~\eqref{I_RD_osc_avg} or \eqref{I_RD_osc_avg_alt}, we can perform the remaining integral numerically.  Because of the factorization, it suffices to compute the integral once for all values of $k$. The result can be written in the form
\begin{align}
    \Omega_\text{GW}(\eta_\text{c}, f) = A^2 Q(n_\text{s}) \left(\frac{k}{k_*}\right)^{2(n_\text{s}-1)} , \label{Omega_GW_RD_power-law}
\end{align}
where $Q(n_\text{s})$ is the numerical coefficient depending on $n_\text{s}$. 
Fig.~\ref{fig:Q} extends Table 1 in Ref.~\cite{Kohri:2018awv} to show the dependence $Q(n_\text{s})$. In the scale-invariant case $n_\text{s}=1$, $Q(n_\text{s}) = 0.822244$. 

The coefficient blows up as $n_\text{s} \to 5/2$. This can be understood as follows. When $\mathcal{P}_\zeta(k)$ is blue tilted, the convergence property of the integral over $t$ and $s$ is determined by the large $t$ behavior. Neglecting the dependence on $s$, $\Omega_\text{GW} \sim \int^\infty \mathrm{d} t \, \frac{(\log t^2)^2}{t^4} t^{2(n_\text{s}-1)}$. This converges for $n_\text{s} < 5/2$ and it diverges toward $n_\text{s} \to 5/2$ as $(5/2 - n_\text{s})^{-3}$.  In the red-tilt case, on the other hand, it is governed by either small $u$ or $v$ limits.  As a rough estimate, if we fix $v = 1$ and take the small $u$ limit $u \ll 1$, the integral over $u$ converges for $n_\text{s} > -2$.  However, we have not established this point without fixing $v=1$ because the convergence of the numerical integral was not good outside the plotting domain of Fig.~\ref{fig:Q}.

\subsubsection{Example 3: Top-hat function\label{sssec:RD_top-hat}}

Next, let us consider the top-hat (or box) function
\begin{align}
    \mathcal{P}_\zeta(k) = \frac{A}{2\Delta} \Theta(k - k_\text{min})  \Theta(k_\text{max} - k), \label{P_zeta_top-hat}
\end{align}
where $A$ is an overall normalization, $k_\text{min/max}$ is the minimum/maximum wave number for a finite value of $\mathcal{P}_\zeta(k)$, and $2\Delta \equiv \ln (k_\text{max}/k_\text{min})$ is the width of the top-hat.  We also use $\widetilde{A} \equiv A / (2 \Delta)$. The top-hat function case was studied in Ref.~\cite{Saito:2009jt} for the first time. We revisit this case here to provide potentially useful methods or formulas.  This example has not been studied in Ref.~\cite{Kohri:2018awv}, so it is a new minor addition. 
The top-hat function is used in the PTA analysis by the NANOGrav collaboration~\cite{NANOGrav:2023hvm}. The approximation of a generic function by multiple top-hat functions is used in Ref.~\cite{LISACosmologyWorkingGroup:2025vdz} to study the prospects of LISA to reconstruct the SIGW models. 

The integration region is cut by the two step functions. The resulting integration region is summarized in Appendix~\ref{sec:region_top-hat}.  

Even for the top-hat function, which looks simple, the integrand of SIGWs in the RD era is too complicated to analytically perform the integral. Examples of numerically obtained GW spectra for the narrow width case $\Delta \ll 1$ are shown by dashed lines in Fig.~\ref{fig:Omega_GW_RD_delta}.  In the following, we focus instead on the case with a sufficiently large $\Delta \gtrsim 1$, i.e., a sufficiently wide top-hat shape. Then, let us first consider the range $k_\text{min} \ll k \ll k_\text{max}$ within the top-hat. In this case, the minimum and maximum of the integral variables $u$ and $v$ are $u_\text{min}= v_\text{min} \ll 1$ and $u_\text{max} = v_\text{max}\gg 1$, meaning that the dominant part of the integrand is inside the integration region~\cite{Saito:2009jt}.  Thus, $\Omega_\text{GW}(k)$ is not sensitive to the far separated scales $k_\text{min}$ or $k_\text{max}$. It should approximately reproduce the scale-invariant case ($k_\text{min} \to 0$ and $k_\text{max} \to \infty$) up to the overall coefficient $\widetilde{A}^{2}$. Then, the only nontrivial parts of the spectrum are the IR part $k \lesssim k_\text{min}$ and the UV part $k\gtrsim k_\text{max}$. Therefore, we consider the two limiting cases $k_\text{max}\to \infty$ (with fixed $k_\text{min}$) and $k_\text{min}\to 0$ (with fixed $k_\text{max}$) to focus on the IR and UV behaviors, respectively. 

\begin{figure}[tbph]
\begin{center}
\subcaptionbox{$k_\text{max}\to \infty$}{\includegraphics[width=0.48 \columnwidth]{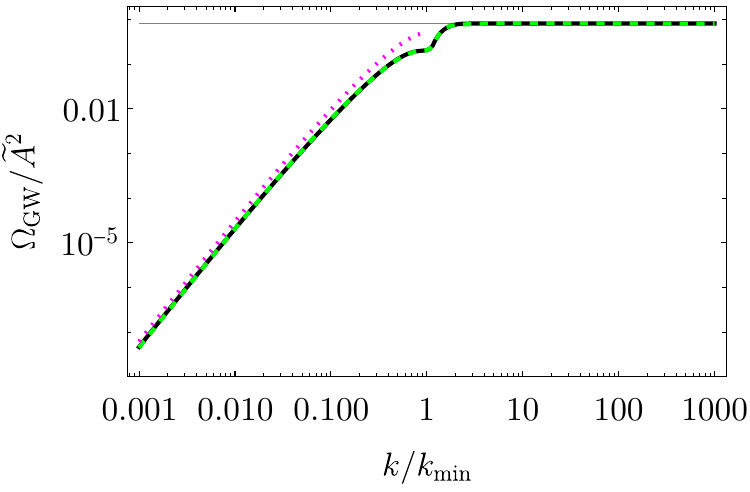}}
\subcaptionbox{$k_\text{min}\to 0$}{\includegraphics[width=0.48 \columnwidth]{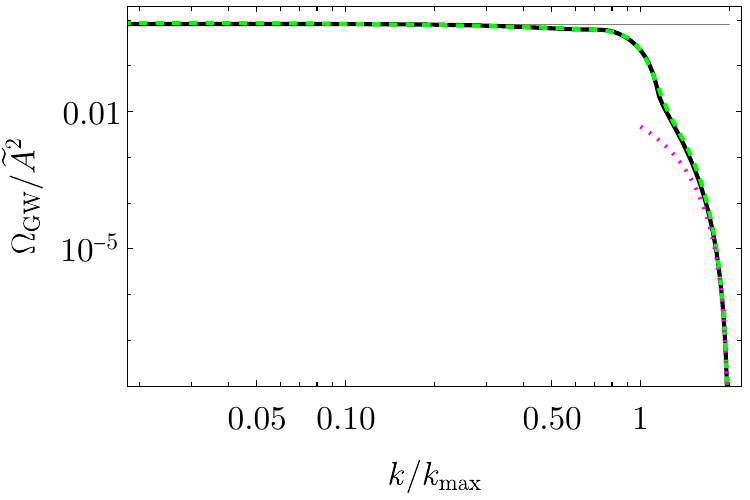}}
\end{center}
\caption{The normalized GW spectra induced by the top-hat $\mathcal{P}_\zeta(k)$. The left and right panels show the limits $k_\text{max} \to \infty$ and $k_\text{min}\to 0$. 
The solid black lines show the numerical results, while the dashed green lines show Pad\'e-like fits [eqs.~\eqref{Omega_GW_RD_top-hat_IR-fit} and \eqref{Omega_GW_RD_top-hat_UV-fit} in Appendix~\ref{sec:fits}]. The thin horizontal gray lines show the scale-invariant case ($k_\text{min} \to 0$ and $k_\text{max} \to \infty$ with fixed $\widetilde{A}$). The dotted magenta lines show the approximations for the IR and UV limits [eqs.~\eqref{Omega_GW_RD_top-hat_IR-limit} and \eqref{Omega_GW_RD_top-hat_UV-limit}].}
\label{fig:Omega_GW_RD_top-hat}
\end{figure}

The result of the numerical integral is shown by the solid black lines in Fig.~\ref{fig:Omega_GW_RD_top-hat}. The left and right panels show the IR and UV behavior, respectively. As expected, the normalized value of $\Omega_\text{GW}$ on the plateau $(k_\text{min}\ll k \ll k_\text{max})$ reproduces that of the scale-invariant case $Q(1) = 0.822244$ (the thin horizontal gray line). 

In the IR limit, the minimum of the integration variable $t$ is $t_\text{min} = 2 k_\text{min}/k -1 \gg 1$, so we can use the large $t$ approximation. Taking the leading term of $t$, one can perform the integral. Then, one can take the leading term in the IR limit $k \to 0$ to obtain
\begin{align}
    \Omega_\text{GW,\,RD}^\text{(top-hat,\,IR limit)}(k)  =\frac{16}{15}\widetilde{A}^{2} \kappa_\text{min}^3 \ln\left( \frac{\kappa_\text{min}}{2}\right)^2, \label{Omega_GW_RD_top-hat_IR-limit}
\end{align}
where $\kappa_\text{min}\equiv k / k_\text{min}$. 
This is shown by the dotted magenta line on the left panel of Fig.~\ref{fig:Omega_GW_RD_top-hat}. 
The cubic power is consistent with the universal causality tail produced in the RD era, and the logarithmic correction is due to the resonant production of the GWs on subhorizon scales.

In the UV part close to the edge of the maximum wavenumber $2 k_\text{max}$, we can make the opposite approximation with small $t \ll 1$. Taking the leading-order of $t$ and neglecting the minor dependence on $s$ inside the log, $\log(3-s^2)\approx \log3$ (remember that $|s|\leq 1$), one can perform the integral.  Then, taking the leading-order term in $(2 - \kappa_\text{max}) \ll 1$ with $\kappa_\text{max}\equiv k / k_\text{max}$, the UV edge of the spectrum can be approximated as 
\begin{align}
    \Omega_\text{GW,\,RD}^{\text{(top-hat,\,UV limit)}} = & 25 \left(1- \mathrm{arctanh} \left(\frac{211}{275}\right)\right)^2 \widetilde{A}^{2}(2 - \kappa_\text{max})^4 \nonumber \\
    \approx & 4.66678 \times 10^{-3} \, \widetilde{A}^{2} (2 - \kappa_\text{max})^4. \label{Omega_GW_RD_top-hat_UV-limit}
\end{align}
This is shown by the dotted magenta line on the right panel of Fig.~\ref{fig:Omega_GW_RD_top-hat}. 

For practical purposes, we consider Pad\'e-like fits in Appendix~\ref{sec:fits}. The dashed green lines in Fig.~\ref{fig:Omega_GW_RD_top-hat} show the fits $\Omega_{\text{GW,\,RD}}^{\text{(top-hat, IR fit)}}$ [eq.~\eqref{Omega_GW_RD_top-hat_IR-fit}] and  $\Omega_{\text{GW,\,RD}}^{\text{(top-hat, UV fit)}}$ [eq.~\eqref{Omega_GW_RD_top-hat_UV-fit}] for the left and right panels, respectively.

\begin{figure}[tbph]
\begin{center}
\includegraphics[width=0.6 \columnwidth]{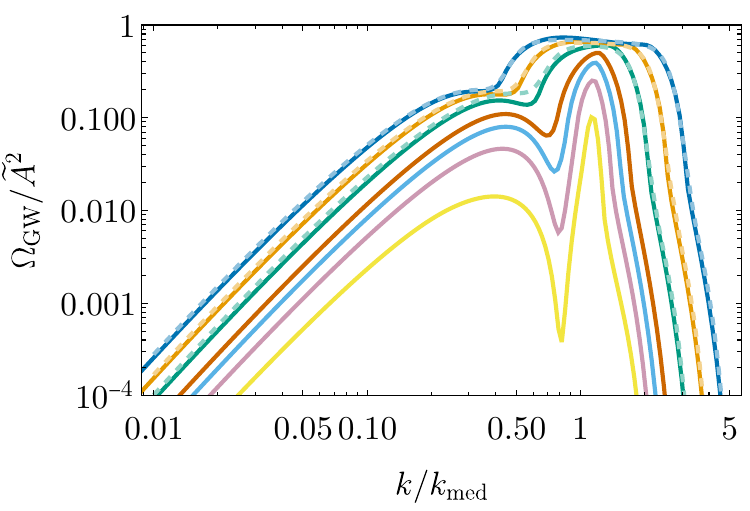}
\end{center}
\caption{Comparison of the numerical results (solid darker lines) of the SIGW spectrum induced in an RD era by the top-hat spectrum of the curvature perturbations with finite widths and their approximations [eq.~\eqref{Omega_GW_RD_top-hat_fit}] (dashed lighter lines). From outside to inside, the width  is $\Delta = 1$ (blue), $0.8$ (orange), $0.6$ (bluish green), $0.4$ (vermilion), $0.3$ (sky blue), $0.2$ (reddish purple), and $0.1$ (yellow).  The fit is better for larger values of $\Delta$. The dashed lines are not plotted for $\Delta \leq 0.4$, where the fit is poor.  The horizontal axis is normalized by $k_\text{med}\equiv \sqrt{k_\text{min}k_\text{max}}$.}
\label{fig:Omega_GW_RD_top-hat_width}
\end{figure}

For a sufficiently large width $\Delta \gtrsim \mathcal{O}(1)$, we can use the following approximation
\begin{align}
    \Omega_{\text{GW,\,RD}}^{\text{(top-hat, fit)}}(\eta_\text{c}, f) = Q(1)^{-1}\widetilde{A}^{-2} \,\Omega_{\text{GW,\,RD}}^{\text{(top-hat, IR fit)}}(\eta_\text{c}, f) \, \Omega_{\text{GW,\,RD}}^{\text{(top-hat, UV fit)}}(\eta_\text{c}, f). \label{Omega_GW_RD_top-hat_fit}
\end{align}
Comparison of this approximation and the numerical result is shown in Fig.~\ref{fig:Omega_GW_RD_top-hat_width}. For $\Delta = 0.6$, we see a clear difference, while the fitting quality is marginal for $\Delta \approx 0.8$. The fit is good, though not perfect, for $\Delta \gtrsim 1$.  

\subsection{Matter-dominated universe\label{ssec:MD}}
In an MD universe, the equation-of-state parameter is $w=0$, the scale factor behaves as $a\propto \eta^2$, and the conformal Hubble parameter satisfies $\mathcal{H}=2/\eta$. The equation of motion for the transfer function $T_\Phi$ becomes $T''_\Phi + (6/\eta) T'_\Phi = 0$. The normalized and regular solution is
\begin{align}
    T_\Phi (k \eta ) = 1.
\end{align}
The source function is a constant too
\begin{align}
    f_\text{MD}(u, v, k \bar \eta) = \frac{6}{5}.
\end{align}
The function $I(u, v, k \eta)$ is 
\begin{align}
    I_\text{MD}(u, v, x) = \frac{6\left(x^3 + 3 x \cos x - 3 \sin x \right)}{5 x^3}, \label{I_MD}
\end{align}
where again we used the notation $x \equiv k \eta$. 
The small $x$ behavior is $3 x^2 /25$, while the large $x$ asymptotic value is $6/5$. 
The late-time limit of the integration kernel is
\begin{align}
    I^2_\text{MD}(u, v, x\gg 1) = \frac{36}{25}. 
\end{align}
Note that this is a constant and $I_\text{MD}(u, v, x)$ does not oscillate at $x\to \infty$. This means that the induced tensor mode does not oscillate like waves.  We have assumed rapid oscillations of GWs on subhorizon scales around eq.~\eqref{Omega_GW_def}, so, strictly speaking, we cannot discuss $\Omega_\text{GW}$ as defined above in the MD universe.\footnote{Alternatively, we can discuss $\Omega_\text{GW}$ in an MD era with some caveats.  Suppose that we are ultimately interested in a more realistic setup where the MD era transitions into an RD era, in which case the frozen tensor mode begins to oscillate.  Along this line, we regard the constant value as the stored potential energy of would-be GWs.  When we introduced the oscillation average, we multiplied the potential energy by a factor of $2$ to take into account the kinetic energy, but it is absent in the MD universe.  To compensate this factor we multiply $1/2$ in the definition of the ``oscillation average'' of $I^2_\text{MD}(u, v, x\gg1 )$,
\begin{align}
\overline{I^2_\text{MD}(u, v, x \gg 1)} =& \frac{18}{25}.
\end{align}
} In this sense, it is clearer to discuss the power spectrum $\mathcal{P}_h(k)$ without the oscillation average. 

In an MD era, density perturbations $\delta = \delta\rho/\rho$ grow as $\delta \propto a$ in a linear regime.  When it becomes nonlinear, $\delta = \mathcal{O}(1)$, the perturbative systems of equations of motion become invalid. Since $\Phi$ is related to $\delta$ by the Poisson equation, $\Phi$  will be affected by the nonlinearity of $\delta$. In this case, the tensor mode induced by $\Phi$ also has uncertainty. 
In the following examples, we consider power spectra $\mathcal{P}_\zeta (k)$ that has a UV cutoff $\mathcal{P}_\zeta (k > k_\text{max})=0$ and consider scales not much below it and time not much after the horizon reentry of the mode $k = k_\text{max}$ to neglect the nonlinearity issue.

\subsubsection{Example 1: Delta function}
Let us consider the delta-function case, eq.~\eqref{P_zeta_delta}. 
The induced tensor spectrum is 
\begin{align}
    \mathcal{P}_h(k)=& \frac{144}{25} \left(\frac{k}{k_*}\right)^{-2} \left( 1 - \left(\frac{k}{2k_*}\right)^2 \right)^2 A^2 \Theta (2 k_* - k). \label{P_h_MD_delta}
\end{align}
This is shown by the dashed black line on the right panel of Fig.~\ref{fig:Omega_GW_MD_width}. 
In contrast to the RD-era case, $\mathcal{P}_h$ does not decay at late time because it is sourced by a square of $\Phi \sim \text{const}.$ This is additional evidence that the induced tensor mode has not yet behaved like usual freely propagating waves decoupled from the source. If we convert it to $\Omega_\text{GW} = \mathcal{P}_h /192  \times (k \eta)^2$ with the caveats discussed above, it has an additional factor of $k^2$ and it grows with time $\propto \eta^2 \propto a$ since the energy density of the sourced tensor perturbations scales as $a^{-2}$ in an MD era while the background energy density scales as $a^{-3}$. 

\subsubsection{Example 2: Top-hat function\label{sssec:MD_top-hat}}
We again consider the top-hat function case, eq.~\eqref{P_zeta_top-hat}, for $\mathcal{P}_\zeta (k)$. This case with a finite $k_\text{min} > 0$ has not been studied in Ref.~\cite{Kohri:2018awv}, and it is an extension of the $k_\text{min}\to0$ limit studied there. 

The integration region in the top-hat case is summarized in Appendix~\ref{sec:region_top-hat}. Depending on the sizes of $k$, $k_\text{min}$, and $k_\text{max}$, the geometric shape varies.  For each case, one can analytically calculate $\mathcal{P}_h (k)$. It can be expressed as follows
\begin{align}
    \mathcal{P}_h(k) =& \mathcal{P}_h(k)|_{k < \min [2 k_\text{min}, k_\text{max} - k_\text{min}]}+ \mathcal{P}_h(k)|_{k_\text{max} - k_\text{min} \leq k < 2 k_\text{min}} \nonumber \\
    & +\mathcal{P}_h(k)|_{2k_\text{min} \leq k < k_\text{max}- k_\text{min}}+ \mathcal{P}_h(k)|_{\max [2 k_\text{min}, k_\text{max}-k_\text{min}] \leq k < k_\text{max}+k_\text{min} }  \nonumber \\
    &+ \mathcal{P}_h(k)|_{k_\text{max}+k_\text{min} \leq k \leq 2 k_\text{max}} .
\end{align}
Note that the second and third terms cannot be simultaneously nonzero.

The explicit expressions of these contributions are given below.
\begin{align}
&   \frac{875}{3} r^2 \kappa_\text{max}^2 \widetilde{A}^{-2}\mathcal{P}_h(k)|_{k < \min [2 k_\text{min}, k_\text{max} - k_\text{min}]}  \nonumber \\
=& 1792(1-r)r^2 \kappa_\text{max} - 560r^2 \kappa_\text{max}^2 - 768 r (1-r) \kappa_\text{max}^3 + 105 (1 + r^2) \kappa_\text{max}^4 , \\
&  \frac{875}{3} r^2 \kappa_\text{max}^2 \widetilde{A}^{-2}\mathcal{P}_h(k)|_{2k_\text{min} \leq k < k_\text{max}- k_\text{min}}  \nonumber \\
=& r^2 \left( 256 r^6 \kappa_\text{max}^{-4} - 896 r^4 \kappa_\text{max}^{-2} + 1792 \kappa_\text{max} -2520 \kappa_\text{max}^2 +768 \kappa_\text{max}^3 + 105 \kappa_\text{max}^4 \right) ,\\
&  \frac{875}{3} r^2 \kappa_\text{max}^2 \widetilde{A}^{-2}\mathcal{P}_h(k)|_{k_\text{max} - k_\text{min} \leq k < 2 k_\text{min}}  \nonumber \\
=& 2 r (1-r)^6 (15+ 26r +15r^2))\kappa_\text{max}^{-4} -56 r (1-r)^4 (3 + 4r +3r^2) \kappa_\text{max}^{-2} + 420 r (1-r^2)^2 \nonumber \\
& -840 r (1-r)^2 \kappa_\text{max}^2 + 105 (1-r)^2 \kappa_\text{max}^4 ,\\
&  \frac{875}{3} r^2 \kappa_\text{max}^2 \widetilde{A}^{-2}\mathcal{P}_h(k)|_{\max [2 k_\text{min}, k_\text{max}-k_\text{min}] \leq k < k_\text{max}+k_\text{min} }  \nonumber \\
=& r \left( \left(30 -128 r + 168 r^2 - 140 r^4 + 168 r^6 + 128 r^7 + 30 r^8\right)\kappa^{-4}_\text{max}  \right.  \nonumber \\
& \quad -56 (3-8r +5 r^2 + 5 r^4 + 8 r^5 + 3 r^6) \kappa_\text{max}^{-2} +420 (1-r^2)^2 + 1792 r^2 \kappa_\text{max} \nonumber \\
& \left. \quad -280 (3 + r + 3r^2) \kappa_\text{max}^2 + 768 \kappa_\text{max}^3 - 105 (2-r) \kappa_\text{max}^4  \right),\\
&  \frac{875}{3} r^2 \kappa_\text{max}^2 \widetilde{A}^{-2}\mathcal{P}_h(k)|_{k_\text{max}+k_\text{min} \leq k \leq 2 k_\text{max}}  \nonumber \\
=& r^2 (1 - 2 \kappa_\text{max}^{-1})^4 \left( -16 - 32 \kappa_\text{max} + 16 \kappa_\text{max}^2 + 72 \kappa_\text{max}^3 + 105 \kappa_\text{max}^4 \right),
\end{align}
where $\kappa_\text{max} \equiv k / k_\text{max}$ and $r \equiv k_\text{min} / k_\text{max}$.  By definition, $0 \leq r \leq 1$, and from the momentum conservation, $0 \leq \kappa_\text{max} \leq 2$.  
In the above expressions,  we have omitted the Heaviside step function for each domain of $k$ for simplicity: \textit{e.g.}, $\mathcal{P}_h(k)|_{k_\text{max}+k_\text{min}\leq k \leq 2 k_\text{max}}$ should be understood as the expression written above times $\Theta(k - (k_\text{max}+k_\text{min})) \Theta(2 k_\text{max}-k)$.

\begin{figure}[tbph]
\begin{center}
\subcaptionbox{}{\includegraphics[width=0.48 \columnwidth]{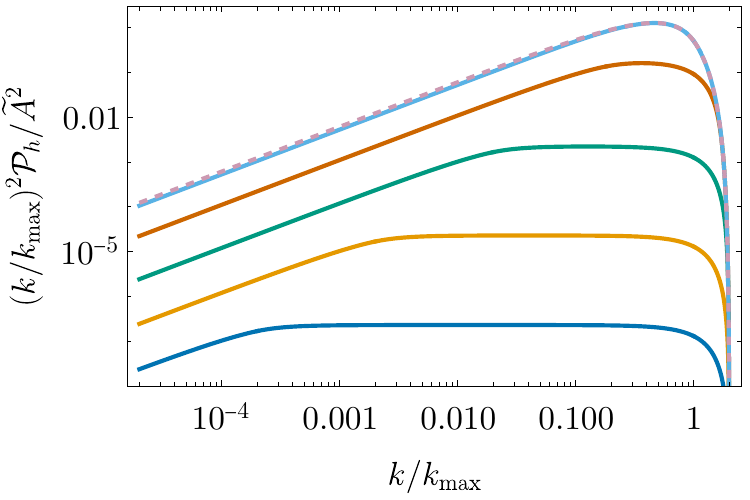}}
\subcaptionbox{}{\includegraphics[width=0.48 \columnwidth]{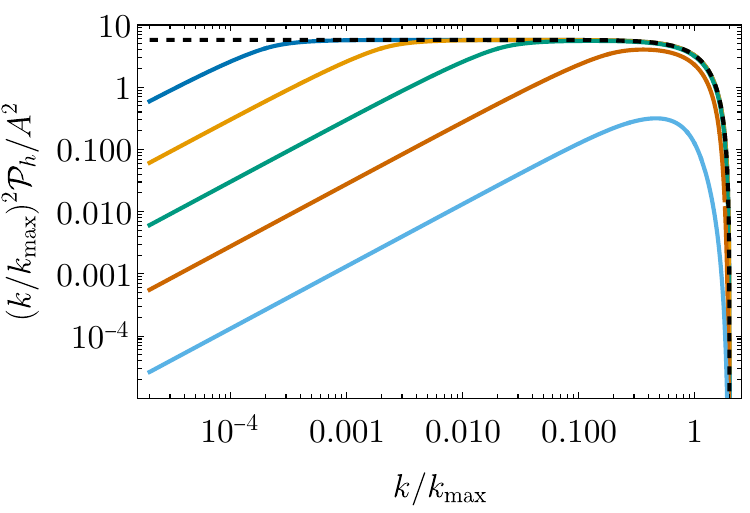}}
\end{center}
\caption{Dependence of the induced $\Omega_\text{GW}$ on the width of the top-hat $\mathcal{P}_\zeta (k)$ in an MD era. The normalizations are different between the left and right panels ($\widetilde{A} = A/(2\Delta)$). The blue, orange, bluish green, vermilion, and sky blue lines correspond to  
$\Delta = 10^{-4}$, $10^{-3}$, $10^{-2}$, $10^{-1}$, and $1$, respectively. 
The dashed reddish-purple line on the left panel is the limit $k_\text{min} \to 0$. The dashed black line on the right panel is the delta-function limit $k_\text{min}/k_\text{max} \to 1$. }
\label{fig:Omega_GW_MD_width}
\end{figure}

Examples of the GW spectra with different choices of the width parameter $\Delta$ are shown in Fig.~\ref{fig:Omega_GW_MD_width}.  
Let us take some limits. The limit $k_\text{min} \to 0$ with fixed $\widetilde{A} (= A/(2\Delta))$ (see the left panel) and dropping the tilde reproduces the case of the scale-invariant spectrum with the UV cutoff in Ref.~\cite{Kohri:2018awv}, \textit{i.e.},
\begin{align}
    \mathcal{P}_h(k) =&\frac{3 \widetilde{A}^2}{875} \times \begin{cases}
        \left(1792 \kappa^{-1}_\text{max} -2520 + 768 \kappa_\text{max} + 105 \kappa_\text{max}^2 \right) & (0 < k \leq k_\text{max}) \\
        \left(1-2 \kappa_\text{max}^{-1} \right)^4 \left( -16 \kappa^{-2}_\text{max} -32 \kappa^{-1}_\text{max} + 16 + 72 \kappa_\text{max} + 105 \kappa_\text{max}^2 \right) & (k_\text{max} < k \leq 2 k_\text{max})
    \end{cases}.
\end{align}
This is shown by the dashed reddish-purple line on the left panel of Fig.~\ref{fig:Omega_GW_MD_width}. 
The leading term for small $\kappa_\text{max}$ reproduces the result of  Ref.~\cite{Assadullahi:2009nf}, and we corrected the numerical factor.  
Another limit is $k_\text{min} \to k_\text{max}$ (the dashed black line on the right panel), which reproduces eq.~\eqref{P_h_MD_delta}. In this limit, only the range $k_\text{max}-k_\text{min} \leq k < 2 k_\text{min}$ remains nontrivial (has finite support).

Other examples of (approximate) analytical formulas for the SIGW spectrum were developed in the literature. The following cases were studied for $\mathcal{P}_\zeta(k)$: a sum of multiple delta functions~\cite{Cai:2019amo}, the lognormal function~\cite{Pi:2020otn}, and the broken power law case~\cite{Li:2024lxx}.

\section{Effects of the transitions between cosmic epochs \label{sec:transitions}}

\subsection{General case\label{ssec:general_transitions}}
Suppose that there are $N$ cosmological epochs and that we can use the instantaneous transition approximation for each transition. Then, we can divide the time integral for $I(u,v,x)$ into $N$ pieces as follows.
\begin{align}
    I(u, v, x) =& \int_0^x \mathrm{d} \bar{x} \frac{a(\bar \eta)}{a(\eta)} k G_k(\eta, \bar{\eta}) f(u, v, \bar x) \nonumber 
    \\
    =& \sum_{i=1}^N \int_{x_{\text{eq},i-1}}^{x_{\text{eq},i}} \mathrm{d}\bar{x} \frac{\bar{a}(\bar \eta)}{a(\eta)} k G^{(i)}_k(\eta, \bar \eta) f^{(i)}(u, v, \bar{x}), \label{I_complete}
\end{align}
where $x_{\text{eq}, i}$ is the $i$-th equality time between the era $i$ and era $i+1$ with $x_{\text{eq}, 0}\equiv 0$ and $x_{\text{eq}, N}\equiv x$, $G_k ^{(i)}(\eta, \bar \eta)$ is Green's function for GWs produced during the $i$-th era, and $f^{(i)}(u, v, \bar x)$ is the function representing the transfer function of the source that produced GWs during the $i$-th era.  We need to keep track of the transitions \emph{before} the $i$-th era for $f^{(i)}(u,v,\bar x)$ and those \emph{after} the $i$-th era for $G_k^{(i)}(\eta, \bar \eta)$. At transitions, we should impose appropriate boundary conditions for $G_k$ and $f$ (or $\Phi$) and their derivatives.  An example is the requirement of continuity of the zeroth and first derivatives at the transition, though it is not always valid as we will see below.

Without further specifying the cosmological history, $G_k^{(i)}$ and $f^{(i)}$ are not determined. However, $T_\Phi$ in the $i$-th era, $T_\Phi^{(i)}$, should be a sum of two independent solutions in the era with the equation-of-state parameter $w^{(i)}$. Similarly, $G_k^{(i)}$ should be a sum of two independent solutions in the era with $w^{(i)}$. We derive the general formula of the integral for unspecified coefficients of these independent solutions.

\subsection{Master formula for a transient RD/MD era\label{ssec:master_formula}}
 Let us now focus on an RD era and an MD era in turn. 
\subsubsection{Master formula for a transient RD era}
 In an RD era, we can express $T_\Phi(\bar x)$ as a linear combination of $j_1 (\bar{x}/\sqrt{3})/\bar x$ and $B y_1(\bar{x} /\sqrt{3})/\bar{x}$ and $k G_k(\eta, \bar \eta)$ as a linear combination of $\sin \bar x$ and  $D \cos \bar x$.  Therefore, we are interested in the following quantity that generalizes $a(\eta) I_\text{RD}(u, v, x)$,
\begin{align}
    \mathcal{I}_\text{RD}(u, v, x_1, x_2) =\int_{x_1}^{x_2} \mathrm{d}\bar x \, \bar{x} \left(C(x)\sin \bar x + D(x) \cos \bar x \right) f_\text{RD}(u, v, \bar{x})|_{T_\Phi(\bar x)= 3\sqrt{3}(A j_1 (\bar{x}/\sqrt{3})+B y_1(\bar{x} /\sqrt{3}))/\bar{x}}. 
\end{align}
The last factor is 
\begin{align}
    &f_\text{RD}(u, v, \bar{x})|_{T_\Phi(\bar x)= 3\sqrt{3}(A j_1 (\bar{x}/\sqrt{3})+B y_1(\bar{x} /\sqrt{3}))/\bar{x}}  \nonumber \\
    &= \frac{1}{u^3 v^3 \bar x^6}\left(E^{\cos, -}\cos \frac{s \bar{x}}{\sqrt{3}} +E^{\sin, -}\sin\frac{s \bar x}{\sqrt{3}} + E^{\cos, +}\cos\frac{(t+1)\bar x}{\sqrt{3}}+ E^{\sin,+} \sin \frac{(t+1)\bar x}{\sqrt{3}} 
 \right),
\end{align}
where we mixed the notation $t=u+v-1$ and $s=v-u$ to slightly compactify the expression, and the factors $E$ are functions of $u$, $v$, and $\bar x$,
\begin{align}
    E^{\cos,\mp}(u, v, \bar{x})=& \pm6 \left(A(u) A(v) \pm B(u) B(v)\right) \left(54 - 6(u^2 + v^2 \mp 3 u v) \bar x^2 + u^2 v^2 \bar x^4 \right) \nonumber \\
    & \qquad \left. \pm 12 \sqrt{3} \left( A(u) B(v) \mp A(v) B(u)\right) (u \mp v) \bar x (\pm 9 + u v \bar x ^2) \right), \\
    E^{\sin,\mp}(u, v, \bar{x})=& \pm 6  \left(A(v) B(u) \mp A(u) B(v)\right) \left( 54 - 6(u^2 + v^2 \mp 3 u v) \bar x^2 + u^2 v^2 \bar x^4 \right) \nonumber \\
    & \qquad  \mp 12 \sqrt{3} \left( A(u) A(v) \pm B(u) B(v)\right) (u \mp v) \bar x ( \pm 9 + u v \bar x ^2) ,
\end{align}
With a straightforward calculation, we obtain the master formula for the (transient) RD era
\begin{align}
    \mathcal{I}_\text{RD}(u,v,x_1,x_2)=& \frac{3}{4 u^3 v^3} \left[ \frac{1}{x^4} \sum_{i=\pm}\sum_{j=\pm} \left( F^{ij}(u,v,x) \cos y^{ij} + G^{ij}(u,v,x) \sin y^{ij}  \right)\right]_{x_1}^{x_2} \nonumber \\
    & + \frac{3(u^2 + v^2 -3)^2}{4 u^3 v^3} \sum_{i=\pm}\sum_{j=\pm}\left[ H^{ij}(u,v,x) \mathrm{Ci}(|y^{ij}|) + I^{ij}(u,v,x) \mathrm{Si}(y^{ij})\right]_{x_1}^{x_2}, \label{master-formula_RD}
\end{align}
where $y^{\pm \pm} = \left(1 \pm \frac{v\pm u}{\sqrt{3}}\right)x$ with the first (second) $\pm$ on the left-hand side corresponds to the first (second) $\pm$ on the right-hand side, respectively. The absolute value on $y^{ij}$ is nontrivial only for $y^{-+}$. The functions $F$, $G$, $H$, and $I$ depend on $u$, $v$, and $x$. $H$ and $I$ are given by
\begin{align}
    H^{-\mp}(u,v,x)= & \pm \left( A(u) A(v)\pm B(u)B(v)\right) D(x) - \left( A(v) B(u) \mp A(u) B(v)\right)C(x) ,\\
    H^{+\mp}(u,v,x) =& \pm \left(A(u)A(v)\pm B(u) B(v) \right)D(x) + \left(A(v)B(u) \mp A(u)B(v)\right)C(x), \\
    I^{-\mp}(u,v,x) =& \pm \left( A(u)A(v) \pm B(u)B(v)\right) C(x) + \left(A(v)B(u)\mp A(u) B(v) \right) D(x), \\
    I^{+\mp}(u,v,x) =& \pm \left( A(u)A(v) \pm B(u) B(v)\right) C(x) - \left(A(v)B(u) \mp A(u) B(v) \right) D(x),
\end{align}
and $F$ and $G$ are given in terms of these by
\begin{align}
    F^{\mp -}(u,v,x) =& \pm I^{\mp -}(u,v,x) \left( 18 \left(\mp 1 + \sqrt{3}(u- v)\right) x +\left(\mp 3 + \sqrt{3}(u-v)\right)\left((u+v)^2-3\right)x^3  \right) \nonumber \\
    & - H^{\mp-}(u,v,x) \left( 54 -3 \left(3 + u^2 + v^2 -6 u v \pm 2 \sqrt{3} (v-u) \right) x^2\right), \\
    F^{\mp+}(u,v,x)=& \mp I^{\mp+}(u,v,x)\left( 18\left( \pm1 + \sqrt{3}(u+v)\right)x + \left( \pm 3 + \sqrt{3}(u+v) \right)\left((u-v)^2-3\right)x^3 \right) \nonumber \\
    &- H^{\mp+}(u,v,x) \left( 54-3 \left( 3+ u^2 + v^2 + 6u v \pm 2 \sqrt{3}(u+v)  \right) x^2\right), \\
    G^{\mp-}(u,v,x) =& \mp H^{\mp-}(u,v,x) \left(18 \left( \mp1 + \sqrt{3}(u-v)\right)x + \left(\mp3 + \sqrt{3}(u-v)\right)\left((u+v)^2 - 3\right)x^3 \right) \nonumber \\
    & - I^{\mp-}(u,v,x) \left( 54 -3\left(3 + u^2 +v^2 -6 u v \pm 2 \sqrt{3}(v-u) \right)x^2 \right), \\
    G^{\mp+}(u,v,x) =& \pm H^{\mp+}(u,v,x) \left(18 \left( \pm 1 + \sqrt{3}(u+v)\right)x +\left(\pm 3 + \sqrt{3}(u+v)\right) \left((u-v)^2 -3\right)x^3 \right) \nonumber \\
    & -I^{\mp+}(u,v,x) \left( 54 - 3 \left( 3+u^2 +v^2 + 6 u v \pm 2 \sqrt{3}(u+v) \right)x^2 \right).
\end{align}
The limit of a pure RD era $x_1 \to 0$ with $A=1$, $B=0$, $C=-\cos x$, and $D=\sin x$, can be taken by noting $\lim_{x_1 \to 0} \mathrm{Ci}(A x_1) - \mathrm{Ci}(B x_1) = \log A - \log B$. It is consistent with eq.~\eqref{I_RD}, \textit{i.e.}, $\lim_{x_1 \to 0} \mathcal{I}_\text{RD}(u,v,x_1, x) = x I_\text{RD}(u, v, x)$. 

\subsubsection{Master formula for a transient MD era}
In an MD era, $T_\Phi(\bar x)$ can be expressed as a sum of the two independent solutions $1$ and $\bar x^{-5}$. The power of the decaying mode is sensitive to a small deviation from the pure MD era, but we anyway neglect the decaying mode. The tensor Green's function $k G_k(\eta, \bar \eta)$ is expressed as a linear combination of $\bar x j_1 (\bar x)$ and $\bar x y_1(\bar x)$. Therefore, we are interested in the following quantity for the MD era,
\begin{align}
    \mathcal{I}_\text{MD}(u, v, x_1, x_2) = \int_{x_1}^{x_2} \mathrm{d}\bar x \, \bar{x}^3 \left( C(x) j_1 (\bar x) + D(x) y_1 (\bar x) \right) f_\text{MD} (u, v, \bar {x} ) |_{T_\Phi (\bar x) = A}, \label{master-formula_MD}
\end{align}
where $j_1$ and $y_1$ are the spherical Bessel functions of the first and second kind, respectively. The last factor is 
\begin{align}
    f_\text{MD}(u, v, \bar x) |_{T_\Phi(\bar x) = A } = \frac{6 A(u) A(v)}{5}.
\end{align}
Finally, the master formula for the (transient) MD era is 
\begin{align}
    \mathcal{I}_\text{MD}(u,v,x_1, x_2) =\frac{6 A(u)A(v)}{5} \left[ C(x) \left((3-x^2)\sin x - 3 x \cos x \right) + D(x) \left((x^2-3)\cos x - 3 x \sin x \right)\right]_{x_1}^{x_2}.
\end{align}
When we take $A=1$, $C= x y_1(x)$, and $D=-x j_1(x)$, it reduces to the pure MD case and is consistent with eq.~\eqref{I_MD}, \textit{i.e}, $\lim_{x_1\to 0}\mathcal{I}_\text{MD}(u,v, x_1, x) = x^2 I_\text{MD}(u,v,x)$.

\subsection{Transition between RD and MD eras\label{ssec:RDMDtransition}}
In this subsection, we consider two scenarios with $N=2$ cosmological eras as simple examples of the above discussions: (1) an MD era transitioning to an RD era and (2) an RD era transitioning to an MD era.  

\subsubsection{MD-to-RD transition\label{sssec:MD2RD}}
As mentioned in the introduction, simple discussions of the MD-to-RD transition in Ref.~\cite{Kohri:2018awv} were significantly refined in Refs.~\cite{Inomata:2019zqy, Inomata:2019ivs}.  Although the main scope of this work does not include the findings in Refs.~\cite{Inomata:2019zqy, Inomata:2019ivs}, we briefly give an overview of these works to explain how the results in Ref.~\cite{Kohri:2018awv} were updated. 

An example of the MD-to-RD transition is the reheating after inflation.  Generically, the inflaton potential around the minimum is quadratic, and the equation of motion is that of nonrelativistic matter after coarse-graining of fast inflaton oscillations.  Inflaton should decay to reheat the Universe, after which the energy density is dominated by radiation. There are other examples of particles or objects that can dominate the energy density of the Universe to lead to a transient MD era: massive  particles, coherent oscillations of scalar fields, and macroscopic objects like black holes, $Q$-balls, and oscillons. 

In the case of perturbative decay of the dominating matter with a constant decay rate $\Gamma$, it turned out that the instantaneous decay approximation is inappropriate since it leads to an overestimate for the induced GWs~\cite{Inomata:2019zqy}.  Decay becomes effective when the Hubble parameter becomes comparable to the decay rate, $H \sim \Gamma$, so there is only a single characteristic time scale.  It means that the transition takes about a Hubble time.  The tensor modes enhanced during the MD era are on subhorizon scales during the transition, so the transition time scale is much longer than the oscillation time scale of the tensor modes.  In other words, the transition is slower than the oscillation time scale of GWs.  During the transition, the energy density of the dominating matter field decays exponentially. This is also true for the density perturbations of the matter.  Importantly, the density perturbations of matter grow during the MD era while those of radiation do not.  Therefore, the density perturbations of matter dominate over those of radiation during the transition even around the equality time for the background matter and background radiation. The gravitational potential $\Phi$ mainly feels the matter density perturbations, which decay exponentially, so $\Phi$ also decays exponentially. The sourced tensor mode also decays exponentially, which is the essential reason why the SIGWs after a gradual transition from an MD era to an RD era are not virtually enhanced~\cite{Inomata:2019zqy}. 

If the transition is rapid, \textit{i.e.}, if the time scale of the change of the equations of state of the Universe is shorter than the GW oscillation time scale, the above suppression effect is negligible. In this case, separation into distinct cosmic eras as in eq.~\eqref{I_complete} involves little ambiguity. After the transition, 
\begin{align}
    I(u,v,x) \simeq & \int_0^{x_\text{R}} \mathrm{d}\bar x\, \left(\frac{x_\text{R}}{ x }\right) \left( \frac{\bar{x}}{x_\text{R}}\right)^2 k G_k^{\text{MD}\to\text{RD}} (\eta, \bar \eta) f_\text{MD} (u, v, \bar x) \nonumber \\
    & + \int_{x_\text{R}}^x \mathrm{d}\bar x \, \left(\frac{\bar x}{x} \right) k G_k^{\text{RD}}(\eta, \bar \eta) f_{\text{MD}\to \text{RD}}(u, v, \bar x), \label{I_MD2RD}
\end{align}
where $x_\text{R}\equiv k \eta_\text{R}$ with $\eta_\text{R}$ denoting the transition (Reheating) time, we approximated $a(\eta)/a(\eta_\text{R}) = x/x_\text{R}$ after the transition,\footnote{
This was improved in Ref.~\cite{Inomata:2019ivs} by using $a(\eta)/a(\eta_\text{R})= 2(\eta / \eta_\text{R}) - 1$.  
} $G_k ^{\text{MD}\to \text{RD}} = G_k^{(1)}$ denotes the Green's function sourced during the MD era and propagating in the subsequent RD era, and $f_{\text{MD}\to \text{RD}} = f^{(2)}$ represents the transfer function of the source term that experienced the transition. 

The main focus in the early literature was the contribution in the first line, \textit{i.e.}, the GWs induced during the MD era.  The MD-to-RD transition was explicitly taken into account as above in Ref.~\cite{Kohri:2018awv}, and we have
\begin{align}
    &\int_0^{x_\text{R}} \mathrm{d}\bar x \, \left(\frac{x_\text{R}}{x}\right) \left( \frac{\bar{x}}{x_\text{R}}\right)^2 k G_k^{\text{MD}\to\text{RD}} (\eta, \bar \eta) f_\text{MD} (u, v, \bar x) \nonumber \\
    =& \frac{3}{5 x x_\text{R}^3} \left(3(2x_\text{R}^2 -1)\cos x - 6 x_\text{R} \sin x + 2 x_\text{R}^4 \cos (x- x_\text{R}) + 4 x_\text{R}^3 \sin (x - x_\text{R}) + 3 \cos (x - 2 x_\text{R}) \right).
\end{align}

However, it was pointed out in Ref.~\cite{Inomata:2019ivs} that the second line in eq.~\eqref{I_MD2RD}, \textit{i.e}, the GW contribution induced after the transition is the dominant contribution for a sufficiently rapid transition.  There are two reasons for the enhancement.  First, there is no time for the source to substantially decay during the rapid transition (even for subhorizon modes).  Second, the subhorizon modes begin to oscillate after the transition rapidly compared to the Hubble scale at the transition time. The combination of these two effects implies fast oscillations of deep subhorizon modes with unsuppressed oscillation amplitudes.  This is an interplay between the transient MD era and the subsequent RD era. Without the MD era, the amplitude of the subhorizon modes decays. Without the RD era, they do not oscillate.   The master formula, eq.~\eqref{master-formula_RD}, was used in Ref.~\cite{Inomata:2019ivs} to derive an approximate analytic formula for the SIGW spectrum in the instantaneous transition case. 

The above enhancement mechanism was dubbed the poltergeist mechanism in Ref.~\cite{Inomata:2020lmk}. Applications of the poltergeist mechanism include a triggeron model with the kinematical blocking effect~\cite{Inomata:2019ivs}, a scenario of rapid transition induced by a first-order phase transition in a dark sector~\cite{Inomata:2019ivs}, simultaneous evaporation of PBHs with a narrow mass/spin spectrum~\cite{Inomata:2020lmk, Domenech:2020ssp, Domenech:2021wkk, Bhaumik:2022pil, Bhaumik:2022zdd, Domenech:2023mqk, Borah:2022vsu, Flores:2023dgp, Papanikolaou:2024kjb}, similar mechanisms for $Q$-balls~\cite{White:2021hwi, Kasuya:2022cko, Kawasaki:2023rfx, Yu:2025jgx} and oscillons~\cite{Lozanov:2022yoy}, and a particular parameter space of a rotating axion model~\cite{Harigaya:2023mhl}. See Ref.~\cite{Inomata:2025wiv} for a review of the poltergeist mechanism. 

Recent development of the enhanced SIGWs in the presence of the MD era includes the interpolation~\cite{Pearce:2023kxp} (see also Ref.~\cite{Pearce:2025ywc}) between the gradual~\cite{Inomata:2019zqy} and instantaneous~\cite{Inomata:2019ivs} transition limits and the effect of the difference in velocity perturbations of matter and radiation~\cite{Kumar:2024hsi}.

\subsubsection{RD-to-MD transition}
In the concordance cosmological model, the $\Lambda$CDM model, the RD era is followed by the MD era (but in this case, see footnote~\ref{fn:MLambda}).  Similarly, A transient early MD era in the nonminimal scenario may be preceded by an early RD era.  Let us focus on the RD-to-MD transition in this subsubsection. The expression for $I(u,v,x)$ now reads
\begin{align}
    I(u, v, x) \simeq & \int_0^{x_\text{eq}} \mathrm{d} \bar x \, \left( \frac{x_\text{eq}}{x}\right)^2 \left( \frac{\bar x}{x_\text{eq}} \right) k G_k ^{\text{RD}\to \text{MD}}(\eta, \bar \eta) f_\text{RD}(u, v, \bar x) \nonumber \\
    & + \int_{x_\text{eq}}^x \mathrm{d} \bar x \, \left( \frac{\bar x}{x} \right)^2 k G_k^\text{MD} (\eta, \bar \eta) f_{\text{RD}\to \text{MD}}(u, v, \bar x), \label{I_RD2MD}
\end{align}
where  $x_\text{eq} \equiv k \eta_\text{eq}$ with $\eta_\text{eq}$ denoting the transition (equality) time, we approximated $a(\eta)\propto \eta$ $(\eta^2)$ in the RD (MD) era, respectively,\footnote{
This can be improved by using the exact solution of the scale factor in the presence of radiation and matter, $a(\eta)/a(\eta_\text{eq}) = (\eta/\eta_*)^2 + 2 (\eta/\eta_*)$ with $\eta_* = \eta_\text{eq}/(\sqrt{2}-1)$.
} $G_k^{\text{RD}\to\text{MD}} = G_k^{(1)}$ denotes the Green's function of GWs sourced during the RD era and propagating in the RD era, and $f_{\text{RD}\to\text{MD}} = f^{(2)}$ represents the transfer function of the source term after the transition. 

The contribution in the first line represents the GWs produced during the RD era, which are diluted during the MD era, so there is no effect of enhancement. In eq.~\eqref{Omega_GW_obs}, we consider this contribution since we are primarily interested in sufficiently short-scale GWs.  On the other hand, the contribution in the second line involves a possible enhancement effect associated with the MD era.  As we have seen in the previous subsubsection, the final intensity of the induced GWs significantly depends on the time scale of the transition from the MD era to the subsequent era.  Let us focus on the second line in the rest of this subsubsection. 

In the RD-to-MD transition, we cannot neglect the isocurvature perturbations (see the right-hand side of eq.~\eqref{EoM_Phi_complete}) because the perturbations of matter grow. In fact, if we neglect the nonadiabatic pressure, any infinitesimal perturbations to the pure MD era would forbid the constant $\Phi$ solution.  As is well known, the constant $\Phi$ solution during the MD era is supported by the isocurvature perturbations.\footnote{Here, the terminology (adiabatic or isocurvature) is not about the initial condition but about the time-dependent quantities.}  The large $k$ behavior of the transfer function $T_\Phi$ can be approximated by (see, e.g., Ref.~\cite{Mukhanov:2005sc})
\begin{align}
    T_\Phi(x \gg x_\text{eq}) = & \frac{\ln (c_1 x_\text{eq})}{(c_2 x_\text{eq})^2} , & 
    c_1 =& \frac{2 e^{\gamma -7/2}}{\sqrt{3}(\sqrt{2}-1)} \approx 0.15, &
    c_2 =& \frac{\sqrt{9/10}}{9(\sqrt{2}-1)} \approx 0.25, \label{T_Phi_RD2MD}
\end{align}
where $\gamma$ is the Euler-Mascheroni constant. 
This suppression originates from the fact that the modes that entered the Hubble horizon during the RD era decay on subhorizon scales. Importantly, this effectively acts as a UV cutoff scale.  The second line of eq.~\eqref{I_RD2MD} for $x \gg x_\text{eq}$ becomes
\begin{align}
    \int_{x_\text{eq}}^x \mathrm{d} \bar x \, \left( \frac{\bar x}{x} \right)^2 k G_k^\text{MD} (\eta, \bar \eta) f_{\text{RD}\to \text{MD}}(u, v, \bar x) \simeq & \frac{6}{5} \frac{\ln (c_1 u x_\text{eq})}{(c_2 u x_\text{eq})^2}\frac{\ln (c_1 v x_\text{eq})}{(c_2 v x_\text{eq})^2}.
\end{align}
If we use a fitting formula for $T_\Phi$ valid not only for $x\gg x_\text{eq}$ but also for $x \lesssim x_\text{eq}$, such as the one used in Ref.~\cite{Kohri:2018awv} or a more precise BBKS formula~\cite{Bardeen:1985tr} used in Ref.~\cite{Inomata:2020lmk}, we can generalize the above expression.

\section{Conclusion \label{sec:conclusion}}

We have derived the analytic formula of the integration kernel $I(u, v, x)$ in the RD era [eq.~\eqref{I_RD}] and its late-time oscillation average [eq.~\eqref{I_RD_osc_avg} or eq.~\eqref{I_RD_osc_avg_alt}]. We have also extended the former to the case of a transient RD era [eq.~\eqref{master-formula_RD}]. These are the main results of Ref.~\cite{Kohri:2018awv}. We have also derived the counterparts in the MD era [eq.~\eqref{I_MD} and its generalization~\eqref{master-formula_MD}].

In this paper, we have added some updates to the results in Ref.~\cite{Kohri:2018awv}. 
The main part of the new contributions is about SIGWs in the case of the top-hat $\mathcal{P}_\zeta (k)$, eq.~\eqref{P_zeta_top-hat}. We have studied it focusing on $\Delta \gtrsim \mathcal{O}(1)$ in the RD era and proposed fitting formulas in Sec.~\ref{sssec:RD_top-hat}.  In the case of the MD era, we have derived the fully analytic formulas for $\mathcal{P}_h(k)$ in Sec.~\ref{sssec:MD_top-hat}.  Other minor updates include a more detailed study on $Q(n_\text{s})$ [Fig.~\ref{fig:Q}] and an alternative expression for $\overline{I_\text{RD}(u,v,x \gg 1)^2}$ [eq.~\eqref{I_RD_osc_avg_ana}]. Sec.~\ref{ssec:general_transitions} and Appendixes~\ref{sec:region_top-hat} and \ref{sec:fits} are also new.

The (semi)analytic formulas for the GWs induced by curvature perturbations are useful, practical tools to accelerate the scientific comparison between theories and observations.  These formulas are one of the solid bases to reveal the mysteries of the early Universe and high-energy physics, including inflation, PBHs, Electroweak vacuum metastability, cosmological equations of state, PTA physics of nanohertz GWs, and so on.  

\section*{Acknowledgment}

We thank Kazunori Kohri for the collaboration in the original work~\cite{Kohri:2018awv}.  
This work was supported by the 34th (FY 2024) Academic research grant (Natural Science) No.~9284 from DAIKO FOUNDATION.

\appendix

\section{Integration region for GWs induced from a top-hat \texorpdfstring{$\mathcal{P}_\zeta (k)$}{Pzeta(k)}\label{sec:region_top-hat}}

\begin{figure}[tbph]
\begin{center}
\includegraphics[width=0.6 \columnwidth]{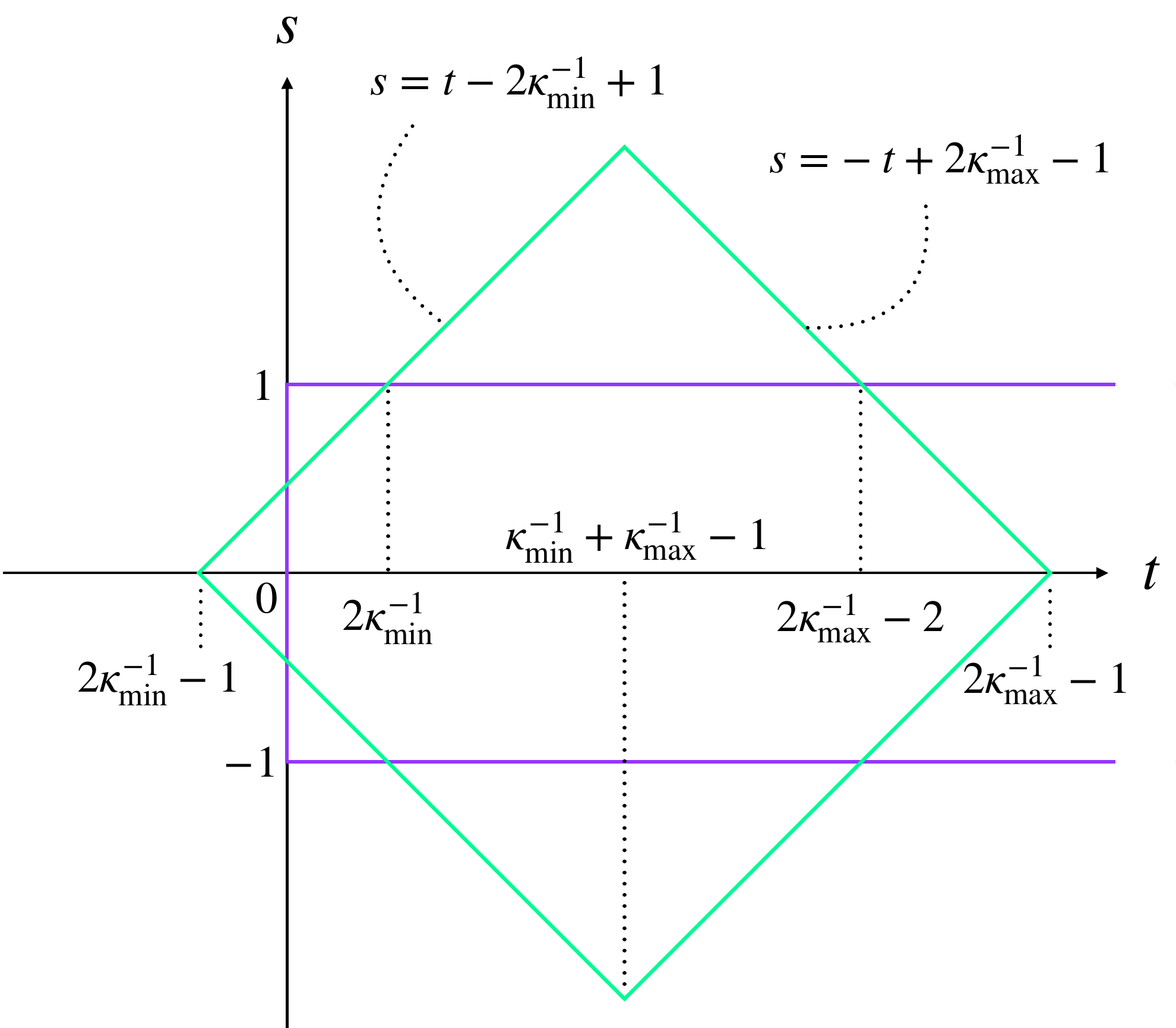}
\end{center}
\caption{Schematic figure showing the integration region. The region enclosed by the purple line shows the original integration domain $t \geq 0$ and $|s|\leq 1$ (remember the symmetry under the sign flip $s \leftrightarrow -s$). The region enclosed by the green line represents the part with nonvanishing $\mathcal{P}_\zeta (u k)$ and $\mathcal{P}_\zeta(v k)$. The overlap between these two regions contributes to the integral for $\mathcal{P}_h$ (or $\Omega_\text{GW}$). Depending on the inequality relation among $k_\text{min}$, $k$, and $k_\text{max}$, the intersection points move, and the geometric shape of the integration domain varies.  In the particular case depicted above, it corresponds to $2 k_\text{min} \leq k < k_\text{max} - k_\text{min}$.}
\label{fig:integration_region}
\end{figure}

Within the integral over $u$ and $v$ for the power spectrum of the induced GWs, the arguments of the power spectrum of the curvature perturbations $\mathcal{P}_\zeta$ are $uk$ and $vk$ (see eq.~\eqref{P_h}).  For the top-hat $\mathcal{P}_\zeta$ case~\eqref{P_zeta_top-hat}, this restricts their range to be within $[k_\text{min}, \, k_\text{max}]$. This leads to the integration rage 
\begin{align}
    \kappa^{-1}_\text{min} \leq u, v \leq \kappa^{-1}_\text{max},
\end{align}
where $\kappa^{-1}_\text{min/max} \equiv k_\text{min/max} / k$, for a fixed wavenumber $k$ of the GWs.
Although the integration cutoffs related to $k_\text{min/max}$ are given directly in terms of $u$ and $v$, the case analysis is complicated. 
It turns out that the case analysis in terms of the variables $t$ and $s$ is more directly related to the final expressions of the SIGW spectrum. 
The geometric shape of the integration region is a polygon whose shape depends on the following cases (see Fig.~\ref{fig:integration_region}).
\begin{enumerate}
    \item $k < \min [2 k_\text{min}, k_\text{max}- k_\text{min}]$
    $$\int_{2 \kappa^{-1}_\text{min}-1}^{2 \kappa^{-1}_\text{min}} \mathrm{d}t \int_0^{t - (2 \kappa^{-1}_\text{min}-1)}\mathrm{d}s + \int_{2 \kappa^{-1}_\text{min}}^{2\kappa^{-1}_\text{max}-2} \mathrm{d}t \int_0^1 \mathrm{d}s + \int_{2 \kappa^{-1}_\text{max}-2}^{2\kappa^{-1}_\text{max}-1}\mathrm{d}t \int_0 ^{2 \kappa^{-1}_\text{max}-1-t}\mathrm{d}s.$$
    \item $2 k_\text{min} \leq k < k_\text{max}-k_\text{min}$
    $$\int_0^{2 \kappa^{-1}_\text{min}} \mathrm{d} t \int_0^{t - (2\kappa^{-1}_\text{min}-1)}\mathrm{d}s + \int_{2 \kappa^{-1}_\text{min}}^{2\kappa^{-1}_\text{max}-2} \mathrm{d}t \int_{0}^{1}\mathrm{d}s  + \int_{2\kappa^{-1}_\text{max}-2}^{2\kappa^{-1}_\text{max}-1} \mathrm{d}t \int_0^{2 \kappa^{-1}_\text{max}-1-t}\mathrm{d}s. $$
    \item $k_\text{max}-k_\text{min} \leq k < 2 k_\text{min}$
    $$\int_0^{\kappa^{-1}_\text{max}+\kappa^{-1}_\text{min}-1}\mathrm{d}t \int_0^{t - (2 \kappa^{-1}_\text{min}-1)}\mathrm{d}s + \int_{\kappa^{-1}_\text{max}+\kappa^{-1}_\text{min}-1}^{2 \kappa^{-1}-1} \mathrm{d}t \int_0^{2\kappa^{-1}_\text{max}-1-t}\mathrm{d}s.$$
    \item $\max[2k_\text{min}, k_\text{max} - k_\text{min} ] \leq k < k_\text{max} + k_\text{min}$
    $$\int_0^{\kappa^{-1}_\text{max}+\kappa^{-1}_\text{min}-1} \mathrm{d} t \int_0^{t - (2\kappa^{-1}_\text{min}-1)}\mathrm{d}s + \int_{\kappa^{-1}_\text{max}+\kappa^{-1}_\text{min}-1}^{2 \kappa^{-1}_\text{max}-1} \mathrm{d}t \int_0^{2\kappa^{-1}_\text{max}-1 - t} \mathrm{d} s.$$ 
    \item $k_\text{max} + k_\text{min} \leq k < 2 k_\text{max}$
    $$\int_0^{2 \kappa^{-1}_\text{max}-1} \mathrm{d}t \int _0^{2 \kappa^{-1}_\text{max}-1 - t}\mathrm{d} s.$$
\end{enumerate}

\section{Pad\'e(-like) approximations for SIGW spectra\label{sec:fits}}
In this appendix, we provide some fitting formulas for GWs induced in the RD era. 

First, let us consider the case of the power law $\mathcal{P}_\zeta (k)$ studied in Sec.~\ref{sssec:RD_power-law}. 
For the $n_\text{s}$ dependence of the coefficient $Q(n_\text{s})$ in eq.~\eqref{Omega_GW_RD_power-law}, we propose a Pad\'e(-like) approximation
\begin{align}
    Q^{\text{(fit)}}(n_\text{s}) = \frac{c_0 + c_1 n_\text{s} + c_2 n_\text{s}^2 + c_3 n_\text{s}^3 + c_4 n_\text{s}^4 + c_5 n_\text{s}^5 + c_6 n_\text{s}^6}{(5-2n_\text{s})^3 + d_4 (5-2n_\text{s})^4 + d_5 (5-2n_\text{s})^5}. \label{Q_fit}
\end{align}
This is shown by the dashed green line in Fig.~\ref{fig:Q} with the parameters in Tab.~\ref{tab:Q_fit}. 
This choice is just an illustration of the idea of the approximation. Both the plotting region in Fig.~\ref{fig:Q} and the fitting region are $-1.395\leq n_\text{s}-1 \leq 1.365$.

\begin{table}[tbhp]
    \centering
    \caption{Parameter values for fitting $Q(n_\text{s})$}
    \label{tab:Q_fit}
    \begin{tabular}{|c|c|c|c|c|}

    \hline 
      $c_0$ & $c_1$ & $c_2$ & $c_3$ & $c_4$  \\ \hline
       $5.02250 \times 10^2$   &  $-8.66221 \times 10^2$   &  $6.44507 \times 10^2$   & $-2.65493 \times 10^2$    & $6.73161 \times 10$  \\ \hline

    \end{tabular}
    
    \begin{tabular}{|c|c|c|c|} \hline
          $c_5$ & $c_6$ & $d_4$ & $d_5$ \\ \hline
          $-1.09928 \times10$    & $9.57800 \times 10^{-1}$    &  $8.41395 \times 10^{-1}$   & $-2.96528\times10^{-2}$ \\ \hline
    \end{tabular}
\end{table}

Next, let us consider the case of the top-hat function $\mathcal{P}_\zeta(k)$ studied in Sec.~\ref{sssec:RD_top-hat}. 
Combining the information of the IR limit~\eqref{Omega_GW_RD_top-hat_IR-limit} and the plateau part asymptoting to the scale-invariant case, we consider the following ansatz,
\begin{align}
\Omega_\text{GW,\,RD}^{\text{(top-hat, IR fit)}}(k) \widetilde{A}^{-2} = \frac{(n_{30}+n_{31}\ln \kappa_\text{min} + n_{32} (\ln \kappa_\text{min})^2 )\kappa_\text{min}^3 }{1 + d_1 \kappa_\text{min} + d_2 \kappa_\text{min}^2 + (d_{30} + d_{31}\ln \kappa_\text{min} + d_{32} (\ln \kappa_\text{min})^2)\kappa_\text{min}^3 }, \label{Omega_GW_RD_top-hat_IR-fit}
\end{align}
where $\kappa_\text{min} \equiv k / k_\text{min}$. 
This is like a Pad\'e approximation, but it is augmented by the logarithmic dependence. 
This is plotted as the dashed green line on the left panel of Fig.~\ref{fig:Omega_GW_RD_top-hat} with the parameter values in Tab.~\ref{tab:IR_fit}. 
These values are just a demonstration of the idea and are not meant as recommended values for precise studies. 

\begin{table}[tbhp]
    \centering
    \caption{Parameter values for fitting the IR part of $\Omega_\text{GW,\,RD}^{\text{(top-hat)}}$.}
    \label{tab:IR_fit}

\begin{tabular}{|c|c|c|} \hline
          $n_{30}$ & $n_{31}$ & $n_{32}$  \\ \hline
          $6.41900\times 10^{-2}$    & $-0.245576$    &  $0.90777$  \\ \hline
    \end{tabular}
    
    \begin{tabular}{|c|c|c|c|c|}
    \hline 
      $d_1$ & $d_2$ & $d_{30}$ & $d_{31}$ & $d_{32}$  \\ \hline
       $ -0.819805$   &  $ -0.881308$   &  $1.00912$   & $-0.898108$    & $1.18961$  \\ \hline

    \end{tabular}

\end{table}

Since the UV limit~\eqref{Omega_GW_RD_top-hat_UV-limit} has quite a limited validity range (see the right panel of Fig.~\ref{fig:Omega_GW_RD_top-hat}), we also provide an approximation formula for the UV part. Since we did not find a good Pad\'e approximation, we consider the following ansatz involving a sum of two terms
\begin{align}
    \Omega_\text{GW,\,RD}^{\text{(top-hat, UV fit)}}(k) \widetilde{A}^{-2} = \frac{c_n  y^4}{1 + c_1 y + c_2 y^2 + c_3 y^3 + c_4 y^4} + \frac{y^4}{d_0 + d_1 y + d_2 y^2 + d_3 y^3 + d_4 y^4}, \label{Omega_GW_RD_top-hat_UV-fit}
\end{align}
with the constraints $c_n + \frac{1}{d_0} = 25(1-\mathrm{arctanh}(211/275))^2$ and $\frac{c_n}{c_4} + \frac{1}{d_4} = Q(1)$, where $y \equiv 2 - \kappa_\text{max}$. 
This is plotted by the dashed green line on the right panel of Fig.~\ref{fig:Omega_GW_RD_top-hat} with the parameter set in Tab.~\ref{tab:UV_fit}. 
Again, the choice is just an illustration of the idea and not recommended for precise studies. 

\begin{table}[tbhp]
    \centering
    \caption{Parameter values for fitting the UV part of $\Omega_\text{GW,\,RD}^{\text{(top-hat)}}$.}
    \label{tab:UV_fit}

\begin{tabular}{|c|c|c|c|c|} \hline
          $c_n$ & $c_1$ & $c_2$ & $c_3$ & $c_4$  \\ \hline
          $-1.10078$    & $ -2.17102$    &  $ 1.57278$ & $ -0.411722$ & $7.74029\times 10^{-2}$ \\ \hline
    \end{tabular}
    
    \begin{tabular}{|c|c|c|c|c|}
    \hline 
      $d_0$ & $d_1$ & $d_2$ & $d_3$ & $d_4$   \\ \hline
       $ 0.904609$   &  $-1.96348$   &  $1.42172$   & $-0.368852$    & $6.64729 \times 10^{-2}$  \\ \hline

    \end{tabular}

\end{table}

\bibliographystyle{utphys}
\bibliography{ref.bib}
\end{document}